\documentclass[aip]{revtex4-1}
\usepackage{graphicx}
\usepackage{subcaption}
\usepackage{amsmath}
\usepackage{epstopdf}
\usepackage{IEEEtrantools}
\usepackage[colorlinks=true,linkcolor=black,citecolor=black]{hyperref}
\newcommand{\widefigurewidth}{0.9\textwidth}
\newcommand{\ud}{\,\mathrm{d}}
\newcommand{\e}{\mathrm{e}}
\renewcommand{\i}{\mathrm{i}}
\newcommand{\sgn}{\mathrm{sgn}}
\newcommand{\citetdot}[1]{\citeauthor{#1}.\citep{#1}}
\newcommand{\citetcomma}[1]{\citeauthor{#1},\citep{#1}}

\draft 
\begin{document}

\title{Experimental validation of the hybrid scattering model of installed jet
noise} 



\author{Benshuai Lyu}
\email[]{bl362@cam.ac.uk}
\author{Ann Dowling}
\affiliation{Department of Engineering, University of Cambridge}

s
\date{\today}
\begin{abstract}
   Jet installation causes jet noise to be amplified significantly at low
   frequencies and its physical mechanism must be understood to develop
   effective aircraft noise reduction strategies. A hybrid semi-empirical
   prediction model has recently been developed based on the
   instability-wave-scattering mechanism. However, its validity and accuracy
   remain to be tested. To do so, in this paper we carry out a systematic
   installed jet-noise experiment in the laboratory using a flat plate instead
   of an aircraft wing. We show that reducing $H$ (the separation distance
   between the flat plate and jet centreline) causes stronger low-frequency
   noise enhancement while resulting in little change to the noise shielding
   and enhancement at high frequencies. Decreasing $L$ (the axial distance
   between the jet exit plane and the trailing edge of the plate) results in
   reduced noise amplification at low frequencies and also weakens both the
   shielding and enhancement at high frequencies. Increasing the jet Mach
   number abates the installation effects. It is shown that the hybrid model
   developed in the earlier work agrees with experimental measurements and can
   capture the effects of varying $H$, $L$ and jet Mach number extremely well.
   It is concluded that the model captures the correct physics and can serve as
   an accurate and robust prediction tool. This new physical understanding
   provides insights into innovative strategies for suppressing installed jet
   noise. 
\end{abstract}


\maketitle 

\section{Introduction}
\label{sec:introduction}
The International Civil Aviation Organization (ICAO), a UN specialized agency,
reports that the global air transportation network doubles every $15$ years,
and this trend is expected to continue by $2030$. Currently, there are more
than $100000$ daily flights in this global network. This existing large fleet
and its foreseeable rapid increase raise concerns for their significant
impact on the environment. One great concern is the noise disturbance that it
causes. Since aircraft noise appeared on the agenda, extensive research has
been carried out to reduce it.

Among the many components of aircraft noise, jet noise still dominates at
take-off. In modern aircraft, however, the aero-engines are installed very
close to the aircraft wings, and jet noise is modified significantly by the
wings and other high lift devices. This modified jet noise is commonly
called installed jet noise, in comparison with the isolated jet noise.
Installed jet noise is significantly louder than isolated jet noise at low
frequencies. To reduce aircraft noise, it is therefore necessary to understand
how and why noise is intensified at these low frequencies in the installed
case.

Research on installed jet noise dates back a few decades and the early research
was mostly experimental work. The low-frequency sound intensification due to
jet installation effects was noted by \citet{Bushell1975} in 1975 in his
full-scale flight tests. A systematic model-scale experimental investigation
was conducted by \citet{Head1976} the following year. They concluded that the
installed noise source has a dipole directivity and that the noise intensity
depends on the sixth power of the jet exit velocity. These installation effects
were further examined and confirmed by the experiments of
\citetcomma{Szewczyk1979} \citet{Way1980} and \citet{Shearin1983} in 1979, 1980
and 1983 respectively.

Noise prediction models and reduction strategies were also attempted in the
early 1980s. The works of \citet{Stevens1983} and \citet{Sengupta1983}
represented two early attempts in developing prediction models for installed
jet noise. The former achieved this by summing up jet noise and core noise
measured for a model-scale aircraft, while the latter proposed
numerically fitted models based also on the data from experimental tests. The
experiment of \citet{Wang1980} investigated the noise reduction feasibility of
using aircraft wings with different acoustic surface properties. However, it
was found that the noise reduction occurred mainly at high frequencies, while
the installed jet noise is mainly relevant at low frequencies.

There were few research activities in the late 1980s and early 1990s. This
completely changed in 1998, when \citet{Mead1998} conducted experiments which
aimed to understand the installation effects for sideline observers. In the
same year, \citet{Bhat1998} proposed an empirically-fitted noise prediction
model. However, attempts to account for the installation were rather empirical.
The work of \citet{Moore2004} was somewhat less so. He/she used a model based
on 3D ray theory to quantify the acoustic propagation effects. A few years
later, \citet{Pastouchenko2007} were among the first to study the jet
installation effects using a Computational Fluid Dynamics (CFD) method. Based
on their numerical results, they claimed it was the downwash of jet mean flow
that caused the noise enhancement.

Research interest has been continuously growing since 2012. This is partly
because the engine bypass ratios and jet diameters continue to increase,
causing the jet installation effects to became increasingly pronounced. While
\citeauthor{Brown2013}'s\citep{Brown2013} measurement added to the experimental
database of installed jet noise, the research focus during this period has been
on developing predictive models. Similar to the approach used by
\citetcomma{Papamoschou2010} \citet{Cavalieri2014} proposed a wave-packet
scattering model, and the far-field sound was calculated using a numerical
Green's function integration and a Boundary Elemment Method (BEM). The same two
modelling strategies were used by \citet{Piantanida2016} to model the installed
jet noise from aircraft wings with swept trailing edges. In both studies, a
good overall agreement for the noise directivity was observed at a few discrete
frequencies. On the other hand, \citet{Vera2015} examined the scattering
problem by a different method -- they used Amiet's approach.\citep{Amiet1976b}
However, the results, as they noted in their paper, were not particularly
convenient for quantifying noise sources because they are not statistically
stationary.

Following \citetcomma{Piantanida2015} the idea of using swept aircraft wings to
reduce installed jet noise was further explored by \citetdot{Nogueira2016} As
an other attempt, \citet{Bastos2017240} investigated the effects of chevron
nozzles on installed jet noise. They found that the installed jet noise was not
sensitive to the chevron geometries when the aircraft wing was close to the
jet, though an overall slight noise reduction could be achieved when they were
away from each other.

In the recent works of the authors,~\citep{Lyu2016b,Lyu2016d} a hybrid
prediction model was developed to account for the installation effects. The
model consists of contributions from both the scattering of Lighthill's
quadrupole sources and the scattering of near-field instability waves by
the trailing edge of the aircraft wing. It is found that the scattering of jet
instability waves accounts for the low-frequency enhancement observed in
experiments. In contrast, at high frequencies, noise is either reduced on the
shielded side or enhanced by around $3$ dB on the reflected side of the plate
in accord with classical acoustic scattering theories. The model agrees well
with the data from early experimental tests.~\citep{Head1976} 

As a continuation of this research, in this paper, we carry out an experimental
investigation. One of the aims of this experimental study is to advance our
understanding of the characteristics of installed jet noise and the near-field
pressure fluctuations (due to jet instability waves), as an aid to developing
noise controlling strategies. The other aim is to further validate the hybrid
prediction model developed in the earlier work of the
authors~\citep{Lyu2016b,Lyu2016d} by obtaining a comprehensive experimental
database. In particular, the hybrid model developed in those works requires the
power spectral densities of the near-field pressure fluctuations as an input.
In experiments, we measure such information specifically for the model. In what
follows, we first briefly review the hybrid model, then describe the
experimental setup. The experimental results then follow. The next section
shows the comparison of the experimental results with model predictions, and
the conclusions of the paper are presented subsequently.

\section{The hybrid prediction model}
\label{sec:theHybridPredictionModel}
\begin{figure}
  \centering
  \includegraphics[width=0.58\textwidth]{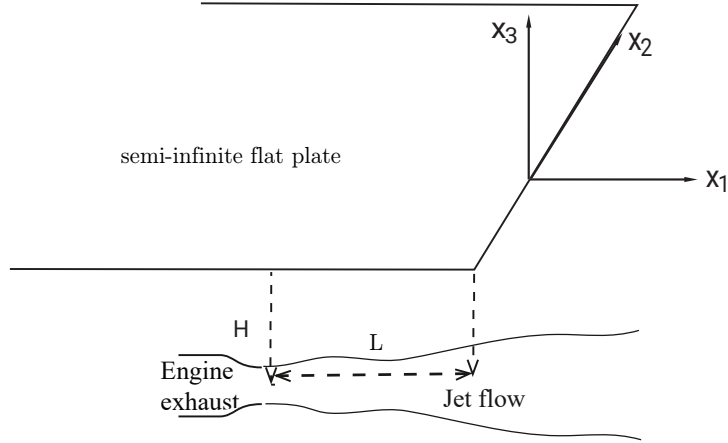}
  \caption{Schematic of the simplified model with a semi-infinite flat plate.}
  \label{fig:jetNearFlatPlate}
\end{figure}

As mentioned in section~\ref{sec:introduction}, the model developed in the
earlier work of the authors~\citep{Lyu2016d} consists of two parts, i.e., the
total sound power spectral density 
\begin{equation}
    \Phi(\omega, \boldsymbol{x}) = \Phi_Q(\omega, \boldsymbol{x}) +
    \Phi_N(\omega, \boldsymbol{x}),
    \label{equ:totalspectrum}
\end{equation}
where $\Phi_Q(\omega, \boldsymbol{x})$ denotes the power spectrum due to
Lighthill's quadrupole sources and $\Phi_N(\omega, \boldsymbol{x})$ denotes the
contribution due to the near-field instability wave scattering. The model
starts by simplifying the wing into a semi-infinite flat plate, as shown in
figure~\ref{fig:jetNearFlatPlate}. The plate is also infinite in the spanwise
direction so that no side-edges are present. $x_1$, $x_2$ and $x_3$ denote the
axes in the streamwise, spanwise and perpendicular directions respectively. $L$
is the horizontal distance between the jet nozzle and the trailing edge of the
flat plate, while $H$ represents the separation distance between the jet
centerline and the bottom surface of the plate. 

\subsection{Quadrupole scattering}
The first part of the model makes use of Lighthill's acoustic analogy theory
using a half-plane scattering Green's function $G(\boldsymbol{x};
\boldsymbol{y}, \omega)$. The details of this function were outlined in an
earlier paper.\cite{Lyu2016d} The far-field power spectral density due to
these quadrupole sources, $\Phi_Q(\boldsymbol{x}, \omega)$, is given by 
\begin{equation}
  \Phi_Q(\boldsymbol{x}, \omega) = c_0^4 \int_{V_y} \int_{V_{\Delta y}}\\
  R_{ijkl}(\boldsymbol{y}, \Delta\boldsymbol{y}, \omega)I_{ijkl}(\boldsymbol{x}, \boldsymbol{y}, \Delta \boldsymbol{y}, \omega) \ud^3\Delta y \ud^3 y,
  \label{equ:PSD}
\end{equation}
where $c_0$ denotes the speed of sound in fluid and
\begin{equation}
  \begin{aligned}
	&R_{ijkl}(\boldsymbol{y}, \Delta\boldsymbol{y}, \omega) =
	\int \overline{T_{ij}(\boldsymbol{y},
	t)T_{kl}(\boldsymbol{y}+\Delta\boldsymbol{y},
    t+\tau)}\frac{\e^{-\i\omega \tau}}{2\pi}\ud \tau, \\
        &I_{ijkl}(\boldsymbol{x}, \boldsymbol{y}, \Delta\boldsymbol{y}, \omega)
	= \frac{\partial^2 G(\boldsymbol{x};\, \boldsymbol{y},
    \omega)}{\partial y_i \partial y_j} \frac{\partial^2
	G^\ast(\boldsymbol{x};\, \boldsymbol{y}+\Delta\boldsymbol{y},
    \omega)}{\partial y_k \partial y_l}.
  \end{aligned}
  \label{equ:tensors}
\end{equation}
$T_{ij}$ in the above equation is Lighthill's stress tensor (see more in
\citet{Lyu2016d}).

The fourth-order space-time correlation function $R_{ijkl}(\boldsymbol{y},
\Delta\boldsymbol{y}, \tau)$ describes the quadrupole sources and can be 
modelled well using a Gaussian function as~\citep{Karabasov2010b}
\begin{IEEEeqnarray}{rCll}
  R_{ijkl}(\boldsymbol{y}, \Delta\boldsymbol{y}, \tau) &=& A_{ijkl}(\boldsymbol{y})\exp \left[ -\frac{|\Delta y_1|}{\bar{u}_1(\boldsymbol{y})\tau_s(\boldsymbol{y})} \right.\nonumber\\
  &&\left.  - \ln2\left(  \left(\frac{\Delta y_1 - \bar{u}_1(\boldsymbol{y})\tau}{l_1(\boldsymbol{y})} \right)^2  +\left(\frac{\Delta y_2}{l_2(\boldsymbol{y})} \right)^2 +\left(\frac{\Delta y_3 }{l_3(\boldsymbol{y})} \right)^2 \right) \right], \IEEEeqnarraynumspace
  \label{equ:correlationFunction}
\end{IEEEeqnarray}
where $A_{ijkl}(\boldsymbol{y}) = C_{ijkl}(2\bar{\rho} k)^2$, $l_i = c_i
k^{3/2}/\epsilon$ and $\tau_s = c_{\tau}k/\epsilon$. Here $\bar{\rho}$ is the
time-averaged fluid density, $\bar{u}_1$ represents the averaged streamwise
velocity, $k$ is the turbulence kinetic energy, $\epsilon$ denotes the
turbulent dissipation rate. These time-averaged quantities can be calculated
using RANS simulation. The constants $C_{ijkl}$, $c_i$ and $c_{\tau}$ can be
obtained by fitting equation~\ref{equ:correlationFunction} to the space-time
correlation data obtained from Large Eddy Simulations (LES). We used the
constants found in the previous work,~\citep{Lyu2016d} i.e. $c_1$, $c_2$,
$c_3$, $c_{\tau}$ are taken to be around $0.4$, $0.23$, $0.23$, $0.3$
respectively, which are close to those obtained by \citetcomma{Karabasov2010b}
but also account for the anisotropy of the turbulence length scales (see
\citet{Mohan2015}). $C_{ijkl}$ remain the same as those found by
\citetdot{Karabasov2010b} Equation~\ref{equ:correlationFunction} is known to be
able to model the fourth-order space-time correlation well, though it does not
account for the cusp behavior at short time
delays.\citep{Karabasov2010b,Mohan2015} Such a model should also suffice for
installed jet noise, provided that the flat plate is not so close to the jet as
to change the flow considerably. Using equation~\ref{equ:correlationFunction},
the cross-spectra is thus obtained by performing the standard Fourier
transformation, which yields
\begin{IEEEeqnarray}{rCl}
  R_{ijkl}(\boldsymbol{y}, \Delta\boldsymbol{y}, \omega) &=&\frac{l_1(\boldsymbol{y})}{2\bar{u}_1(\boldsymbol{y})\sqrt{\pi\ln2}} A_{ijkl}(\boldsymbol{y}) 
  \exp\left[-\frac{l_1(\boldsymbol{y})^2\omega^2}{4\bar{u}_1^2(\boldsymbol{y})\ln2}\right]   \nonumber  \\
  && \exp\left[-\frac{|\Delta y_1|}{\bar{u}_1(\boldsymbol{y})\tau_s(\boldsymbol{y})}  - \i\frac{\omega}{\bar{u}_1(\boldsymbol{y})}\Delta y_1 -\ln2\left( \left(\frac{\Delta y_2}{l_2(\boldsymbol{y})} \right)^2 +\left(\frac{\Delta y_3 }{l_3(\boldsymbol{y})} \right)^2 \right)\right]. \IEEEeqnarraynumspace
  \label{equ:crossSpectrum}
\end{IEEEeqnarray}

The tensor $I_{ijkl}(\boldsymbol{x}, \boldsymbol{y}, \Delta\boldsymbol{y},
\omega)$ in equation~\ref{equ:tensors} depends solely on the Green's function.
Consequently, substituting the free-space Green's function or the one
accounting for half-plane scattering into equation~\ref{equ:PSD} yields results
for an isolated jet or installed jet respectively. Note that this part of the
scattering model does not include the refraction effects of the jet mean flow
on the quadrupole sources. This is partially because we wish to obtain a
Green's function of an analytical function, and including the shear-flow would
make this unlikely to be possible. Detailed discussion on this can be found in
the original paper.\citep{Lyu2016d}

\subsection{Instability scattering}
The second part of the model accounts for the near-field instability
scattering. The contribution of this mechanism to the far-field sound power
spectrum, $\Phi_N(\omega, \boldsymbol{x})$, is given by~\citep{Lyu2016d}
\begin{widetext}
\begin{IEEEeqnarray}{l}
    \Phi_N(\omega, \boldsymbol{x}) = \frac{1}{\pi} \left[\frac{\omega
    x_3}{c_0 S_0^2}\right]^2\sum_{m = -N}^N \left|\frac{\Gamma(c, \mu|_{k_2
	= k\frac{x_2}{S_0}}, \mu_A)}{\mu_A}\right|^2 \Pi(\omega, m)\nonumber \\ 
  \times \left\{\sum_{k = 0}^{[\frac{|m|}{2}]} C_{|m|}^{2k} H^{- 2k + \frac{1}{2}} \gamma_c^{-|m|} \frac{d^{2k}}{dk_2^{2k}}\left[(\gamma_c^2 + k_2^2)^{\frac{1}{2}|m| - \frac{1}{4}} K_{|m| - \frac{1}{2}}\left(H\sqrt{\gamma_c^2 + k_2^2}\right) \right] \right. - \IEEEnonumber \\
  \left. \sgn(m) \sum_{k = 0}^{[\frac{|m|-1}{2}]}  C_{|m|}^{2k+1}  H^{- 2k + \frac{1}{2}}
  \gamma_c^{-|m|} \frac{d^{2k}}{d k_2^{2k}} \left[ k_2 (\gamma_c^2 +
  k_2^2)^{\frac{1}{2}|m|- \frac{3}{4}} K_{|m| - \frac{3}{2}} \left(H
  \sqrt{\gamma_c^2 + k_2^2}\right) \right] \vphantom{\sum_{k =
  0}^{[\frac{|m|-1}{2}]}}\right\}_{k_2 = \frac{kx_2}{S0}}^2, \IEEEeqnarraynumspace
  \label{equ:PhiFinal}
\end{IEEEeqnarray}
\end{widetext}
where $c$ denotes the chord length of the plate, ($x_1, x_2, x_3$) denotes the
Cartesian coordinates of the observer location, the stretched distance of $S_0
= \sqrt{x_1^2 + \beta^2(x_2^2 + x_3^2)}$, $N$ is the largest number of
azimuthal modes that we need to include, $\Pi(\omega, m)$ denotes the $m$-th
near-field pressure power spectral density, the convective radial decay rate is
given by $\gamma_c = \sqrt{(k_1 \beta^2 + k M)^2 - k^2}/\beta$, $K_i$ denotes
the $i$-th modified Bessel function of the second kind, $\sgn(m)$ is the sign
function, $[x]$ denotes the integer not larger than $x$, $C_{m}^{n}$ represents
the binomial coefficient and $\Gamma$ is defined by 
\begin{equation}
    \begin{aligned}
  \Gamma(x, \mu, \mu_A) &=  \e^{\i\mu_A x} E_0(\mu x) \\
  &-\sqrt{\frac{\mu}{\mu-\mu_A}}E_0\left[ (\mu-\mu_A) x \right] - \frac{1}{1 + \i}\e^{\i \mu_A x},
  \end{aligned}
  \label{equ:Definitions}
\end{equation}
where 
\begin{equation}
  \begin{aligned}
      &\mu = k_1 + \sqrt{k^2-k_2^2\beta^2}/\beta^2 + kM/\beta^2, \\
      & \mu_A = k_1 + \frac{k}{\beta^2}(M - \frac{x_1}{S_0}),\\ 
      &E_0(x) = \int_{0}^{x} \frac{\e^{-\i t}}{\sqrt{2\pi t}} \ud t.
  \end{aligned}
  \label{equ:mu}
\end{equation}
Note that although figure~\ref{fig:jetNearFlatPlate} shows a semi-infinite flat
plate, the near-field scattering model can partially capture the effects of the
finite dimensions of the plate using Amiet's approach.\citep{Amiet1976b} Hence
$\Phi_N(\omega,\boldsymbol{x})$ also depends on $c$ as shown in
equation~\ref{equ:PhiFinal}. In the definition of equations~\ref{equ:PhiFinal}
and \ref{equ:mu}, $M$ denotes the Mach number of the ambient flow while $\beta
= \sqrt{1-M^2}$, $k$ (not to be confused with the turbulence kinetic energy)
denotes the acoustic wavenumber while $k_1$ and $k_2$ denote the hydrodynamic
wavenumbers of the near-field pressure in the streamwise and spanwise
directions respectively. 

Equation~\ref{equ:PhiFinal} is the generic form of near-field scattering model.
However, further simplifications can be made in practical cases. For example,
if we assume that the fluctuation is symmetric with respect to the mode number
$m$, i.e. $\Pi(\omega, m) = \Pi(\omega, -m)$, and let $\Pi_s(\omega, m) =
\Pi(\omega, m) + \Pi(\omega, -m)$ for $m \ne 0$, and only two modes need to be
kept, we can show the far-field sound spectral density in the mid-span plane
$(x_2 = 0)$ is 

\begin{IEEEeqnarray}{rCl}
    &&\Phi_N(\omega, \boldsymbol{x}) \approx \left[\frac{\omega x_3}{c_0
    S_0^2}\right]^2 \left\{\left|\frac{\Gamma(c, \mu, \mu_A)}{\mu_A}\right|^2
    \frac{\e^{-2H\gamma_c}}{2\gamma_c^2}\right.\nonumber \\
    &&\left.\times \left(\frac{\Pi_0(\omega, 0)}{K_0^2(\gamma_c r_0)}+ \frac{\Pi_0(\omega, 1)}{K_1^2(\gamma_c r_0)}\right)\right\}_{k_2 = 0, U_c = \overline{U}_c(\omega)},  \IEEEeqnarraynumspace
    \label{equ:PhiMidSpanSemifinal}
\end{IEEEeqnarray}
where $\Pi_0(\omega, m)$ is the $m$-th single-sided spectrum of the incident
near-field evanescent instability waves measured at $r = r_0$ and $U_c$ is the
convection velocity of the near-field instability waves. To ensure acoustic
fluctuations are negligible, one can choose $r_0$ to be small such that the
microphone is sufficiently close to the jet (but not too close to touch the jet
plume). Both $\Pi_0(\omega, 0)$ and $\Pi_0(\omega, 1)$ can be obtained from
experiments or LES. In this study, we use experimental results. Since these
spectra vary with axial position, it makes more sense to put the microphone at
the position where the trailing edge of the flat plate would be if a flat plate
were present, i.e., the place where the incident wave is to be scattered. Also
used in this paper is the frequency-dependent convection velocity
$\overline{U}_c(\omega)$ obtained from LES studies,\cite{Lyu2016d} whose
validity will be verified in section~\ref{sec:comparisonWithPredictionModel}.

\section{Experimental setup}
\label{sec:experimentalSetup} 
\begin{figure}[!hbtp]
    \centering
    \includegraphics{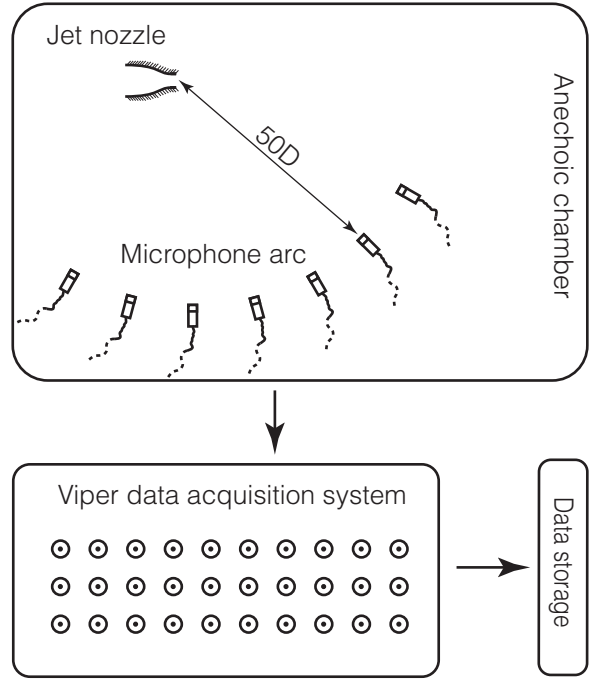}
    \caption{Schematic illustration of the experimental setup.}
    \label{fig:expSetup}
\end{figure}

\begin{figure}[!hbtp]
  \centering
  \begin{subfigure}[b]{0.49\textwidth}
      \centering
      \includegraphics[width=0.9\linewidth]{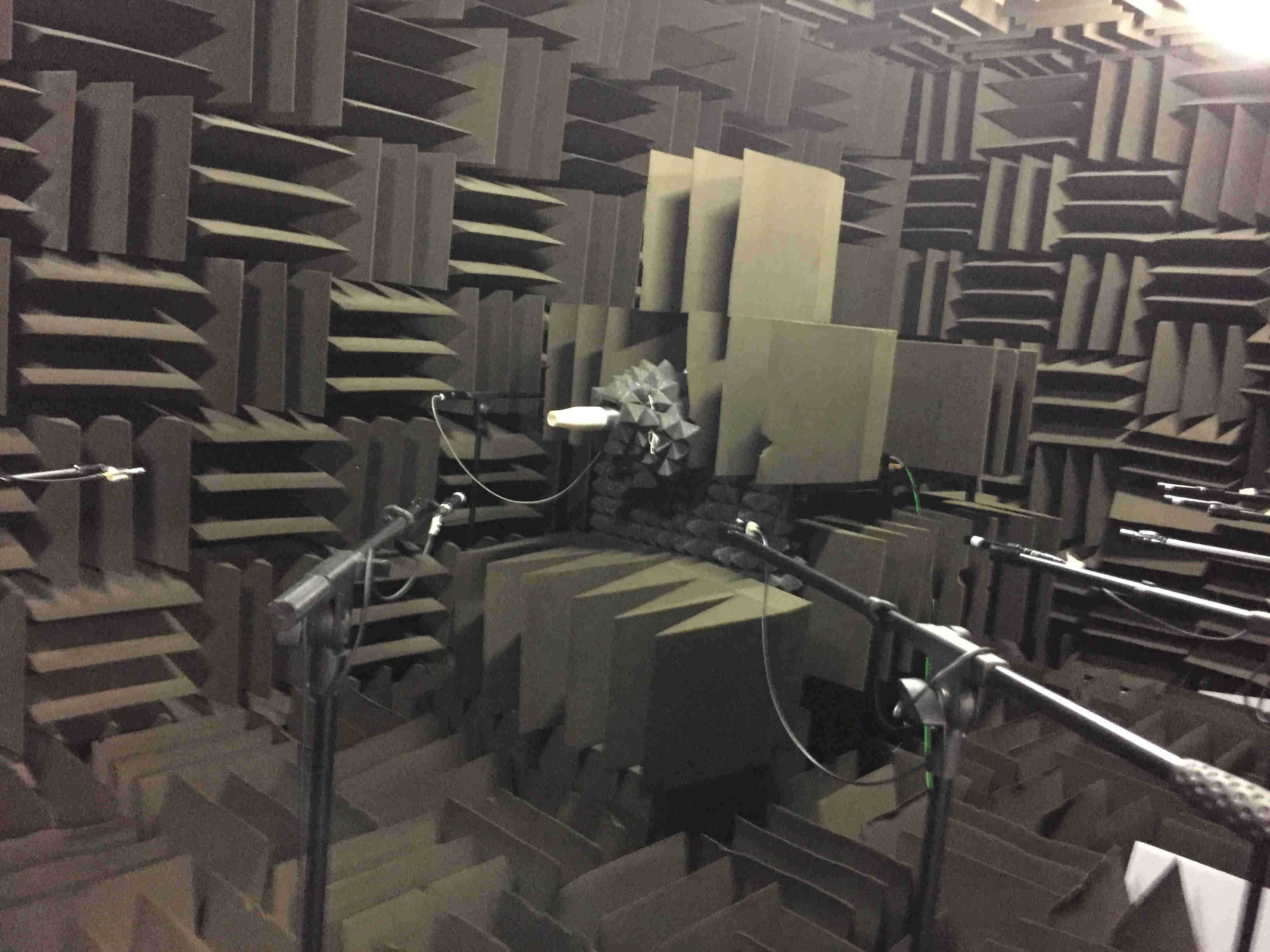}
\caption{The isolated jet noise experiment setup: microphones are located at 50D to the jet nozzle centre, with observer angle ranging from $30^\circ$ to $120^\circ$ to the jet centreline.}
  \label{fig:jetNoiseRig}
  \end{subfigure}
  \begin{subfigure}[b]{0.49\textwidth}
      \centering
      \includegraphics[width=0.9\linewidth]{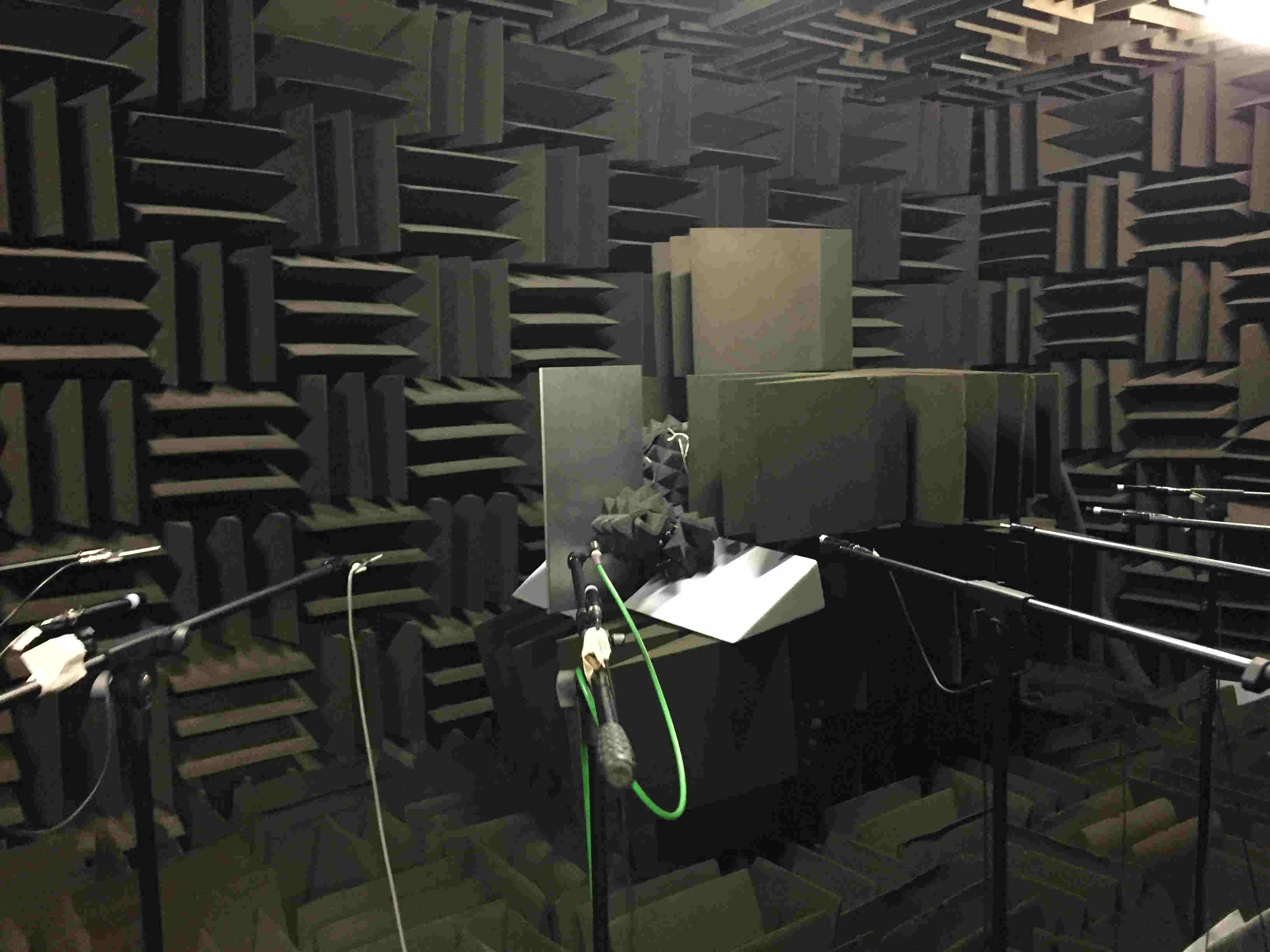}
  \caption{The installed jet noise experiment setup: microphones are located at
      50D to the jet nozzle centre on the shielded side, with observer angle
      ranging from $30^\circ$ to $120^\circ$ to the jet centreline.}
  \label{fig:jetPlateShied}
  \end{subfigure}
  \caption{Experimental setup for the isolated and installed jet noise
  experiments.}
\end{figure}

\begin{figure}
  \centering
  \includegraphics[width=0.28\textwidth, angle=-90]{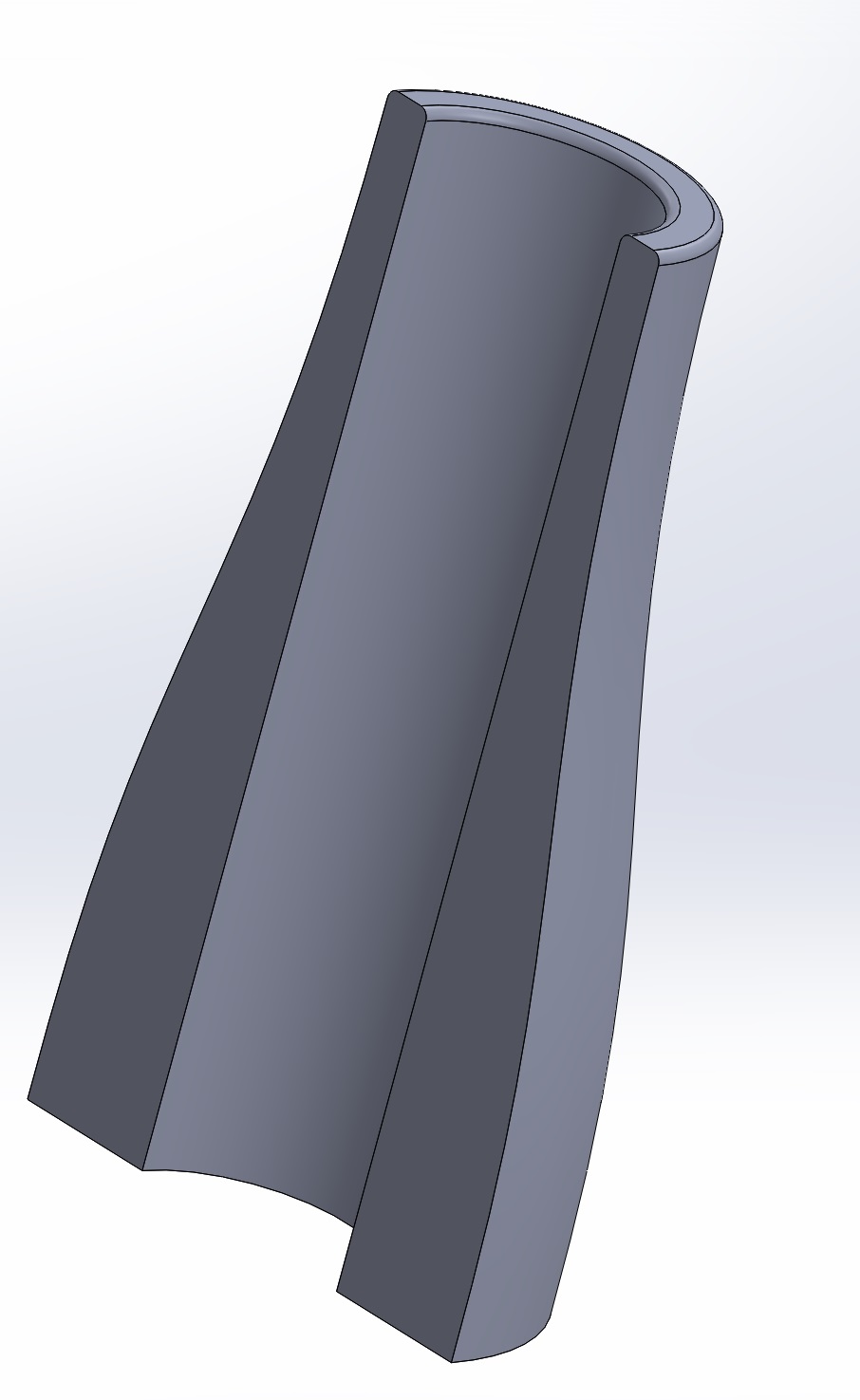}
  \caption{The reference round nozzle used in the experiment with a diameter of
  $2.54\textrm{ cm}$.}
  \label{fig:roundNozzle}
\end{figure}

The schematic illustration of the experiment is shown in
figure~\ref{fig:expSetup}. The experimental rig is placed inside the anechoic
chamber at the Engineering Department of the University of Cambridge, as shown
in figures~\ref{fig:jetNoiseRig} and \ref{fig:jetPlateShied}. The jet nozzle
has a diameter $D = 2.54\textrm{ cm}$. The chamber has a lowest operation
frequency of around 200 Hz. 
As shown in figure~\ref{fig:expSetup}, 7 GRAS 46BE microphones are placed at
$50D$ to the centre of the jet nozzle, at angles in the range of $\theta =
30^\circ$ and $120^\circ$ to the jet centreline. These microphones have a flat
frequency response curve up to $80$ kHz. The electrical signals from these 7
microphones are conditioned, amplified, and then digitalized at a sampling
frequency of $120$ kHz simultaneously using the VIPER data acquisition system
from the IMC Ltd. The jet Mach number is defined by $M_0 = U_j / c_0$, where
$U_j$ is the average jet exit velocity calculated from mass flow rate using an
orifice plate device, and $c_0$, as defined in
section~\ref{sec:theHybridPredictionModel}, is the ambient speed of sound. To
facilitate fast and flexible manufacturing, the round nozzle is 3D printed with
a resolution of $0.1$~mm. It is because of this that the nozzle lip, as can be
seen from figure~\ref{fig:roundNozzle}, has an uncharacteristically large wall
thickness. Data for isolated jet noise is recorded first as reference.

To study the installation effects, a flat plate of $12D \times 24D$ is
subsequently placed nearby the jet, as shown in figure~\ref{fig:jetPlateShied}.
The trailing edge of the plate is at $L$ downstream from the jet nozzle, and
the separation distance between the jet and the plate is $H$, as defined in
figure~\ref{fig:jetNearFlatPlate}. To obtain a comprehensive database on jet
installation effects, both $H$ and $L$ will be varied systematically. The test
matrix is shown in table~\ref{tab:HEffects}. As already mentioned, these tests
are designed not only to study the effects of varying $H$ and $L$ on installed
jet noise, but also to provide a further validation of the hybrid
model~\citep{Lyu2016d} for a comprehensive array of plate positions. 

\begin{table}[!hbtp]
    \centering
\begin{tabular}{cccc}
    \hline
    Test No./Configuration & Mach number& $H$ & $L$\\
    \hline
    1 & $0.5$ & $3D$ & $6D$ \\
    2 & $0.5$ & $2D$ & $6D$ \\
    3 & $0.5$ & $1.5D$ & $6D$ \\
    4 & $0.5$ & $2D$ & $4D$ \\
    5 & $0.5$ & $1.5D$ & $4D$ \\
    6 & $0.5$ & $1.25D$ & $4D$ \\
    \hline
\end{tabular}
\caption{Test parameters for studying the effects of varying $H$ and $L$ using
a $2.54\textrm{ cm}$ round nozzle.}
\label{tab:HEffects}
\end{table}

\begin{table}[!htbp]
    \centering
\begin{tabular}{cccc}
    \hline
    Test No./Configuration & Mach number& $H$ & $L$\\
    \hline
    7 & $0.7$ & $3D$ & $6D$ \\
    8 & $0.7$ & $2D$ & $6D$ \\
    9 & $0.7$ & $1.5D$ & $6D$ \\
    10 & $0.7$ & $2D$ & $4D$ \\
    11 & $0.7$ & $1.5D$ & $4D$ \\
    12 & $0.7$ & $1.25D$ & $4D$ \\
    \hline
\end{tabular}
\caption{Test parameters for studying the effects of varying Mach numbers on
    jet installation effects using a $2.54\textrm{ cm}$ round nozzle.}
\label{tab:MachEffects}
\end{table}

To study the effects of Mach number on installed jet noise, the jet is
subsequently operated at a higher Mach number at $M_0 = 0.7$. Similar to the
tests at Mach number $0.5$, the plate positions are varied systematically. The
complete test matrix is shown in table~\ref{tab:MachEffects}. One can see that
the only difference from that shown in table~\ref{tab:HEffects} is the Mach
number. This is designed such that the Mach number effects can be studied at
each of the many plate positions.

\section{Comparison to other experimental data}
Before discussing experimental results, it is worth comparing the current
experimental data with others' published in the open literature to make sure
the experimental rig is set up properly. It is sufficient to only compare the
reference isolated jet noise spectra. We choose to compare with the data
obtained by Tanna~\citep{Tanna1977b} for a cold $M_0 = 0.5$ jet. The nozzle
used in Tanna's experiment had a diameter of $5.08\textrm{ cm}$, therefore
implying a Reynolds number twice as large as that in this experiment. However,
since both Reynolds number are in the order of $10^5 \sim 10^6$, one can expect
the difference of jet noise spectra caused by different Reynolds numbers
between the two experiments to be insignificant.
\begin{figure}[!hbtp]
  \centering
  \begin{subfigure}[b]{0.49\textwidth}
      \includegraphics[width=\linewidth]{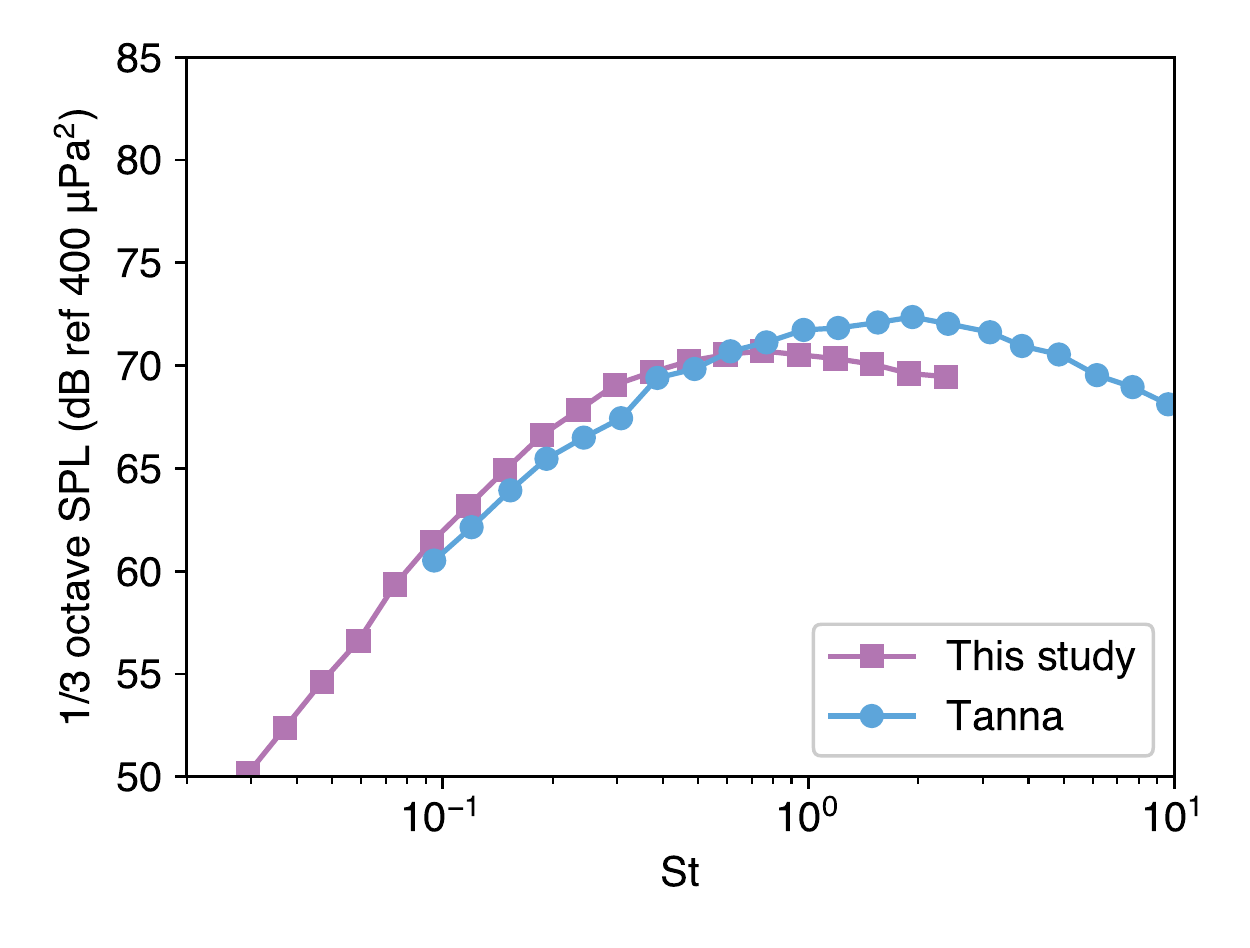}
\caption{$\theta = 90^\circ$}
  \label{fig:compareLyuAndTanna90}
  \end{subfigure}
  \begin{subfigure}[b]{0.49\textwidth}
      \includegraphics[width=\linewidth]{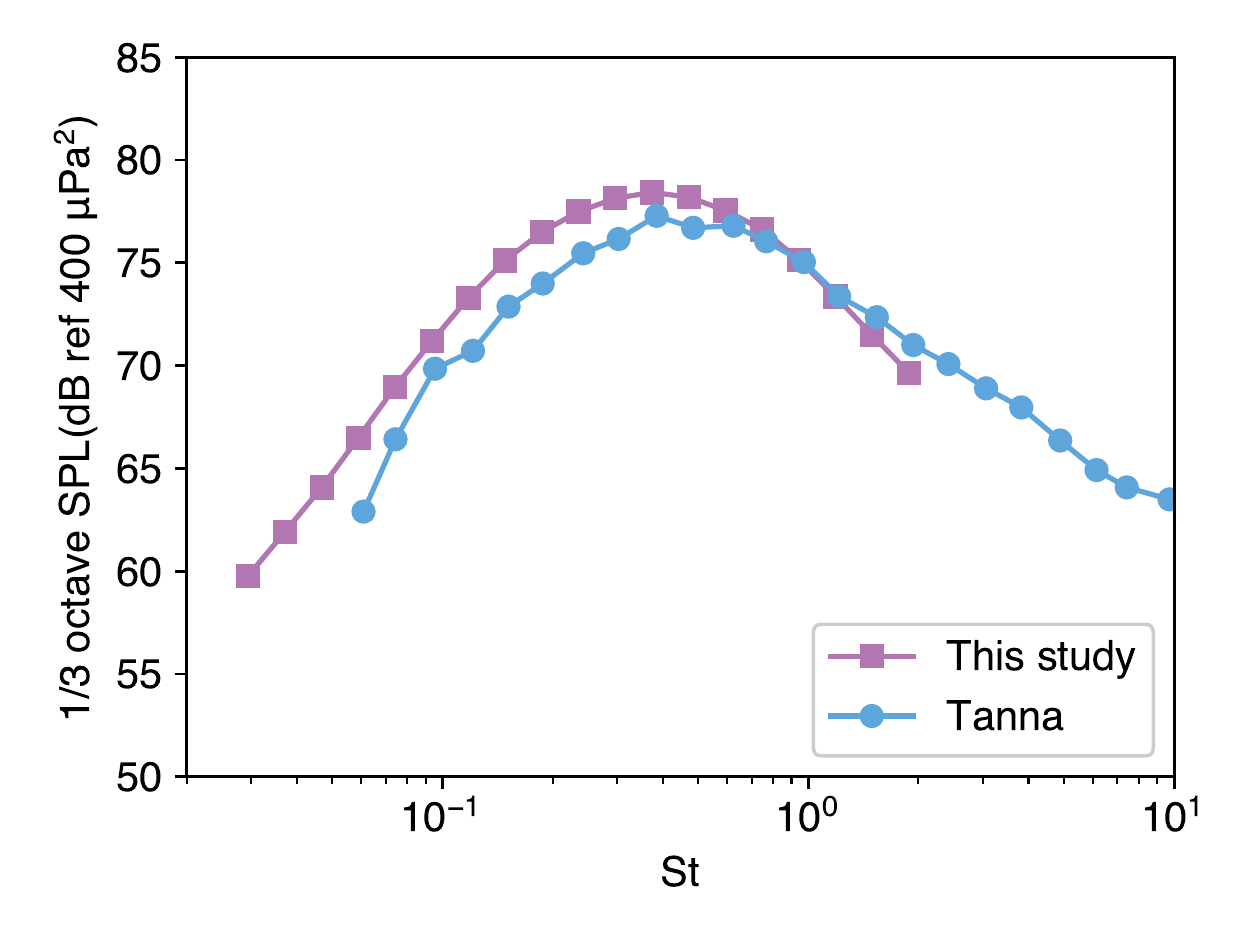}
  \caption{$\theta = 30^\circ$}
  \label{fig:compareLyuAndTanna30}
  \end{subfigure}
\caption{Comparison of isolated jet noise spectra in $1/3$ octaves with
    Tanna's. The microphones are placed at $50D$ from the centre of
nozzle exit.}
\end{figure}
Comparison of the noise spectra at $90^\circ$ and $30^\circ$ to the jet axis
are shown in figures~\ref{fig:compareLyuAndTanna90} and
\ref{fig:compareLyuAndTanna30}, respectively. In both figures, the Strouhal
number is defined as $St = f D/ U_j$. As can be seen, the far-field sound
spectra at $90^\circ$ agree with each other well in their spectral shapes and
absolute magnitudes. There is, however, a small deviation at high frequencies.
This might be due to the different nozzle shapes and inlet flow conditions. The
comparison for the spectra at $30^\circ$ to the jet axis shows a slightly
better agreement, in particular at high frequencies. Considering the many
inevitable differences between the two experiments, such an agreement is
sufficient to show that the experiment is set up properly and the measurement
is reliable. 

\section{Jet noise spectra}
\label{sec:jetNoiseSpectra}
\label{sec:results} 
In this section, we present the jet noise spectra at different observer angles
when the flat plate is placed at different locations. To facilitate a
direct comparison, both isolated and installed jet noise spectra for various
plate positions (table~\ref{tab:HEffects}) are shown in the same figures. 

\subsection{Mach $0.5$ jet}
\begin{figure*}[!htbp]
    \centering
    \includegraphics[width=\widefigurewidth]{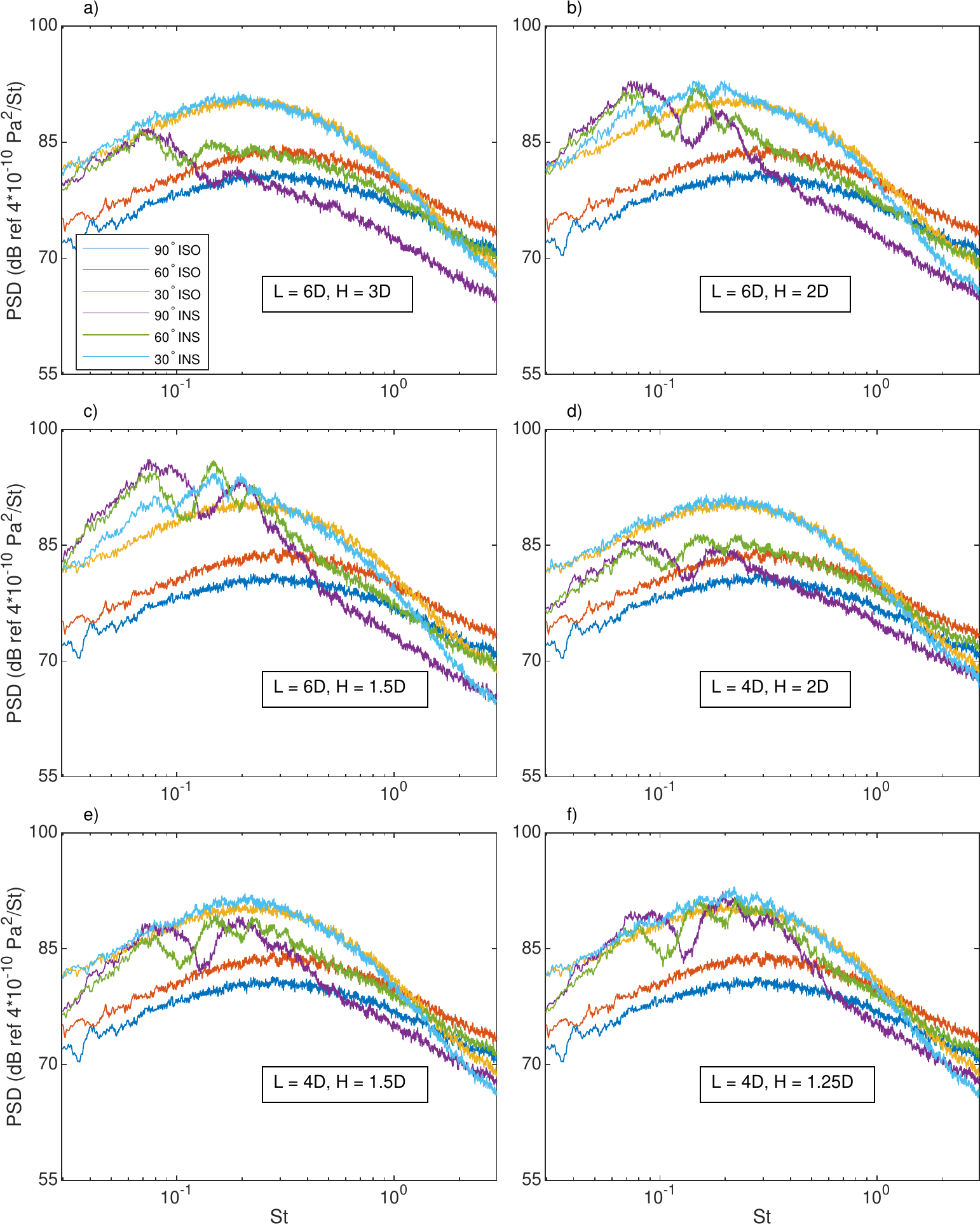}
    \caption{Isolated and installed noise spectra of a Mach number $0.5$ jet at
    $90^\circ$, $60^\circ$ and $30^\circ$ to the jet centreline on the shielded
side, for various plate positions shown inside each figure.}
    \label{fig:RN01_ISOnINS_Mach05}
\end{figure*}

\begin{figure*}[!htbp]
    \centering
    \includegraphics[width=\widefigurewidth]{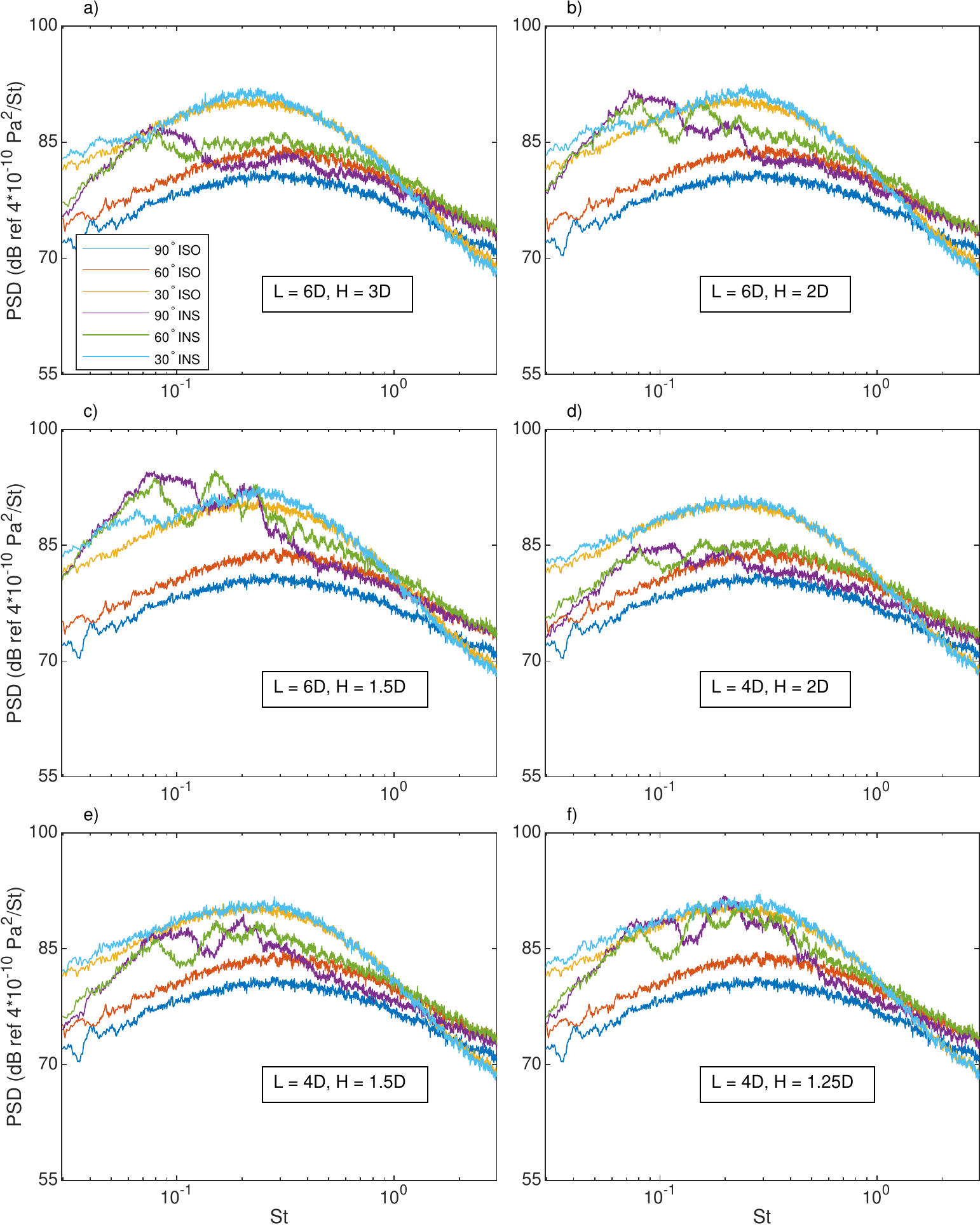}
    \caption{Isolated and installed noise spectra of a Mach number $0.5$ jet at
    $90^\circ$, $60^\circ$ and $30^\circ$ to the jet centreline on the
reflected side, for various plate positions shown inside each figure.}
    \label{fig:Reflected_RN01_ISOnINS_Mach05}
\end{figure*}

We present the results for $M_0 = 0.5$ first. Far-field noise spectra on the
shielded and reflected sides are shown in figures~\ref{fig:RN01_ISOnINS_Mach05}
and \ref{fig:Reflected_RN01_ISOnINS_Mach05}, respectively.
Figure~\ref{fig:RN01_ISOnINS_Mach05}(a-c) shows the far-field noise power
spectra on the shielded side when the plate's edge is at $6D$ downstream from
the jet nozzle but at different radial positions. Both isolated and installed
jet noise spectra are shown. In figure~\ref{fig:RN01_ISOnINS_Mach05}(a), one
can clearly see that the plate enhances the lower-frequency jet noise by up to
10 dB. This noise enhancement is most pronounced at an angle close to
$90^\circ$ to the jet centre line. The noise enhancement at $30^\circ$,
however, is virtually negligible. This is consistent with the earlier
findings~\citep{Head1976, Wang1980, Mead1998, Bondarenko2012, Brown2013,
Amiet1976b, Roger2005} and the edge-scattering mechanism proposed in the
earlier paper~\citep{Lyu2016d} (at low frequencies, the scattered sound has a
dipolar, rather than a cardioid directivity pattern). In the intermediate- and
high-frequency range, jet noise is effectively shielded by the flat plate at
angles close to $90^\circ$ to the jet axis, while these shielding effects
diminish at lower observer angles. Figure~\ref{fig:RN01_ISOnINS_Mach05}(b)
shows the spectra when the plate is at a closer distance to the jet at $H =
2D$. It can be seen that the noise enhancement at low frequencies is now more
pronounced, increasing to more than 15 dB. Also, it can be observed that the
noise increase also occurs in the intermediate frequency range at high observer
angles. An observable noise increase occurs at $30^\circ$ to the jet axis as
well. In the high-frequency regime, one can see that the plate still serves as
an effective noise shield at high observer angles, but little has changed for
the shielding effects compared to figure~\ref{fig:RN01_ISOnINS_Mach05}(a).
Moving the plate closer to $H =1.5D$, as shown in
figure~\ref{fig:RN01_ISOnINS_Mach05}(c), results in an even stronger noise
enhancement. The peak noise increase is now up to 20 dB at $90^\circ$. At such
a close proximity to the plate, jet noise at $90^\circ$ is larger than that at
$30^\circ$. The frequency range of noise increase becomes even wider, leading
to louder noise in the intermediate range. In
figure~\ref{fig:RN01_ISOnINS_Mach05}(a-c), it is shown that the enhanced noise
spectra at low frequencies exhibit a clear oscillation pattern. This
oscillation is more marked when the plate is closer to the jet. A detailed
discussion on this is provided in Appendix A.

The widening of the enhanced frequency range is thought to be closely related
to the radial decay rates of the instability waves. As describe in
section~\ref{sec:theHybridPredictionModel}, when the ambient flow is static,
the decay rate of the instability wave has a dominant exponential form similar
to $\exp(-\gamma_c r)$,\citep{Jordan2013} where
\begin{equation}
    \gamma_c = \sqrt{k_1^2 - k^2}.
    \label{equ:decayRateNoAmbientFlow}
\end{equation}
Therefore, as the frequency increases, $\gamma_c$ increases. This implies that
high-frequency instability waves decay more quickly as the radial coordinate
increases. Consequently, if we place the plate sufficiently far-away from the
jet, only low-frequency instability waves can be seen and scattered into sound
by the flat plate. On the other hand, if the plate is moved closer to the jet,
high-frequency instability waves become stronger, leading to an enhanced
scattering noise at higher frequencies.


The measured noise spectra when the plate is at $L = 4D$ are shown in
figure~\ref{fig:RN01_ISOnINS_Mach05}(d-f). The observer is still on the
shielded side of the plate. Figure~\ref{fig:RN01_ISOnINS_Mach05}(d) presents
the results when $H = 2D$. At such a distance, the noise amplification is
around 8 dB with an observable increase up to $2$ kHz ($St \approx 0.3$). Noise
increase is, as expected, occurring only at high observer angles. At high
frequencies, the noise is reduced by around 3 dB by the shielding effects.
Moving the plate closer to the jet clearly causes a stronger noise increase, as
shown in figure~\ref{fig:RN01_ISOnINS_Mach05}(e), where $H = 1.5D$. At such a
distance, the maximum noise amplification observed is around 12 dB at
$90^\circ$ to the jet centre line. However, there is little noise increase at
$30^\circ$. When the plate is placed at $H = 1.25D$, the tendency of stronger
noise enhancement at a closer distance between the plate and the jet continues.
A noise increase up to 16 dB is observed. The affected frequency range
continues to be wider. The shielding effects remain roughly the same.

The effects of varying $H$ on the installed spectra on the shielded side are
clearly demonstrated by comparing figures~\ref{fig:RN01_ISOnINS_Mach05}(a) to
\ref{fig:RN01_ISOnINS_Mach05}(c), and figures~\ref{fig:RN01_ISOnINS_Mach05}(d)
to \ref{fig:RN01_ISOnINS_Mach05}(f). The effects of varying $L$ at fixed $H$
can be revealed by comparing figures~\ref{fig:RN01_ISOnINS_Mach05}(b) and
\ref{fig:RN01_ISOnINS_Mach05}(d), and figures~\ref{fig:RN01_ISOnINS_Mach05}(c)
and \ref{fig:RN01_ISOnINS_Mach05}(e). When the plate is placed at $H = 2D$, the
noise amplification at $L = 6D$ is clearly much more pronounced than that at $L
= 4D$, with the former one being up to $15$ dB and the latter one being up to 8 dB.
The affected frequency range remains roughly the same. However, it can be seen
that the shielding effects are much more effective when the plate is located
further downstream. Another difference is the observable noise enhancement at
$L = 6D$ but negligible noise increase at $L = 4D$ at $30^\circ$ to the jet
centre line. When the plate is placed at $ H = 1.5D$, the same tendency
remains, while both configurations result in louder noise because of a closer
distance to the jet.

Experimental results measured on the other side of the plate, i.e. the
reflected side, are presented in
figure~\ref{fig:Reflected_RN01_ISOnINS_Mach05}.
Figure~\ref{fig:Reflected_RN01_ISOnINS_Mach05}(a-f) shows the noise spectra
when the plate is at the same positions as those shown in
figure~\ref{fig:RN01_ISOnINS_Mach05}(a-f). At low frequencies, a significant
noise increase is observed, resembling the behaviour on the shielded side, see
figures~\ref{fig:Reflected_RN01_ISOnINS_Mach05}(a) to
\ref{fig:Reflected_RN01_ISOnINS_Mach05}(f). Moreover, comparing
figures~\ref{fig:Reflected_RN01_ISOnINS_Mach05}(a) to
\ref{fig:Reflected_RN01_ISOnINS_Mach05}(c) (and
figures~\ref{fig:Reflected_RN01_ISOnINS_Mach05}(d) to
\ref{fig:Reflected_RN01_ISOnINS_Mach05}(f)) reveals the same behaviour of
louder noise for smaller $H$, when $L$ is fixed. The enhanced sound spectra
appear to be nearly identical to those shown in
figure~\ref{fig:RN01_ISOnINS_Mach05}. However, a careful examination
shows they are not. The most important difference occurs around the lowest
frequencies. For example, comparing figures~\ref{fig:RN01_ISOnINS_Mach05}(a) and
\ref{fig:Reflected_RN01_ISOnINS_Mach05}(a) indicates that the noise increase is
somewhat less pronounced at $90^\circ$ and $60^\circ$. On first thought, this
would invalidate the proposed near-field scattering mechanism, which entails a
perfectly symmetric noise radiation across the plate. 
\begin{figure*}[!hbtp]
    \centering \includegraphics[width=0.8\textwidth]{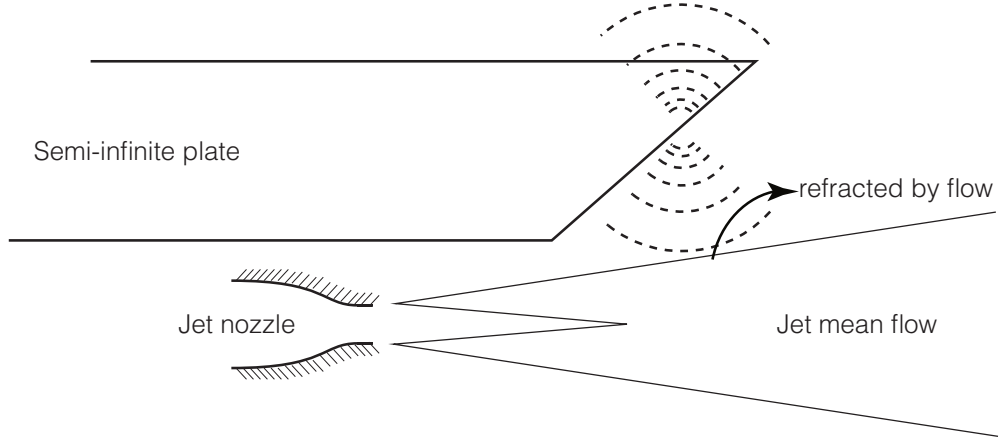}
    \caption{The schematic diagram illustrating the refraction effects of the
    jet mean flow on the reflected side of the flat plate.}
    \label{fig:jetBlockage}
\end{figure*}
However, revisiting the full problem suggests that this is caused by the
refraction effects of the jet mean flow on the reflected side. To show this
more clearly, a schematic diagram is presented in figure~\ref{fig:jetBlockage}.
As can be seen, the scattered sound originates near the edge and propagates
both above and below the edge. However, the sound has to pass through the jet
plume in order to reach an observer placed in the far-field on the reflected
side. This is different on the other side, where no jet plume is present.
Therefore, the discrepancies between the observed spectra on both sides are due
to the jet refraction effects. Decreasing the value of $L$, as can be seen by
comparing figures~\ref{fig:Reflected_RN01_ISOnINS_Mach05}(b) and
\ref{fig:Reflected_RN01_ISOnINS_Mach05}(d) (or
figures~\ref{fig:Reflected_RN01_ISOnINS_Mach05}(c) and
\ref{fig:Reflected_RN01_ISOnINS_Mach05}(e)), causes the noise increase to be
less significant.

However, the installation effects in the high-frequency regime are considerably
different from those on the shielded side. As suggested by its name, noise
increase is observed for spectra on the reflected side due to sound reflection
at high frequencies by the plate.
Figure~\ref{fig:Reflected_RN01_ISOnINS_Mach05}(a) shows that the high-frequency
noise increase is most pronounced at $90^\circ$, and less so at small observer
angles (small $\theta$). Comparing
figures~\ref{fig:Reflected_RN01_ISOnINS_Mach05}(a) to
\ref{fig:Reflected_RN01_ISOnINS_Mach05}(c) (and
figures~\ref{fig:Reflected_RN01_ISOnINS_Mach05}(d) to
\ref{fig:Reflected_RN01_ISOnINS_Mach05}(e)), one can see that changing $H$
while $L$ is fixed does not significantly change this reflection-caused noise
increase at all observer angles, in contrast to the low-frequency noise
enhancement. On the other hand, shortening $L$ causes
high-frequency reflection to be less notable. For example, one can find, by
comparing figures~\ref{fig:Reflected_RN01_ISOnINS_Mach05}(b) and
\ref{fig:Reflected_RN01_ISOnINS_Mach05}(d), that noise increase at high
frequencies is up to $3$ dB when $L = 6D$ while only half is achieved
when $L = 4D$ (at $90^\circ$).

In summary, installed jets exhibit a significant noise increase at low
frequencies compared to isolated jets. The noise spectra have slightly
non-symmetric dipolar directivity patterns due to the asymmetry caused by jet
refraction. At high frequencies, jet noise is noticeably suppressed due to the
plate shielding effects on the shielded side and slightly (e.g., around $2 - 3$
dB) enhanced on the reflected side. Decreasing $H$ while $L$ is fixed results
in a stronger noise increase at low frequencies. At high frequencies, on the
other hand, it causes little change to both the shielding and reflecting
effects. Decreasing $L$ while $H$ is kept constant results in less significant
noise increase at low frequencies. In addition, at high frequencies, both the
shielding and reflecting effects become noticeably less effective.

\subsection{Mach $0.7$ jet}
\begin{figure*}[!htbp]
    \centering
    \includegraphics[width=\widefigurewidth]{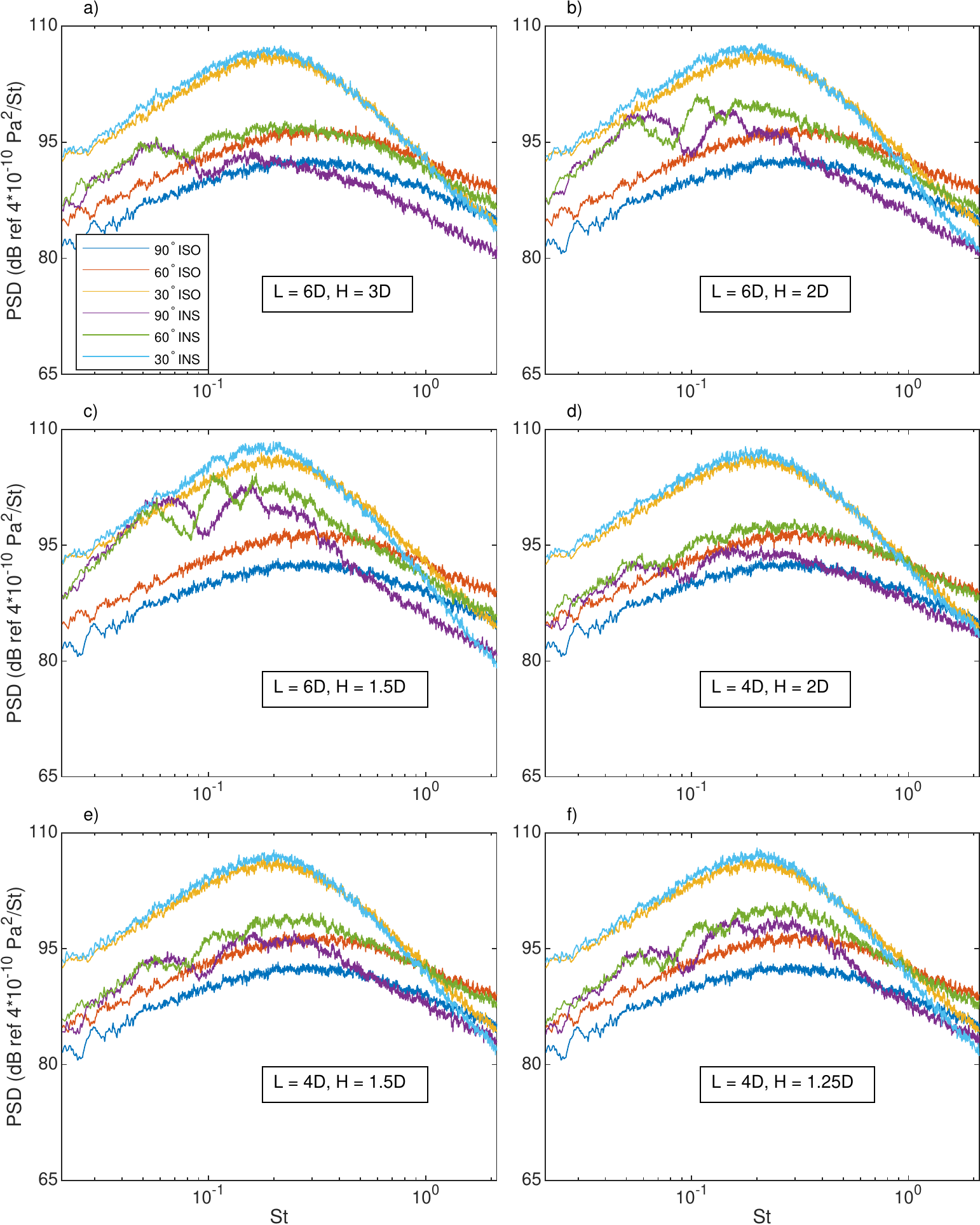}
    \caption{Isolated and installed noise spectra of a Mach number $0.7$ jet at
    $90^\circ$, $60^\circ$ and $30^\circ$ to the jet centreline on the shielded
side, for various plate positions shown inside each figure.}
    \label{fig:RN01_ISOnINS_Mach07}
\end{figure*}

\begin{figure*}[!htbp]
    \centering
    \includegraphics[width=\widefigurewidth]{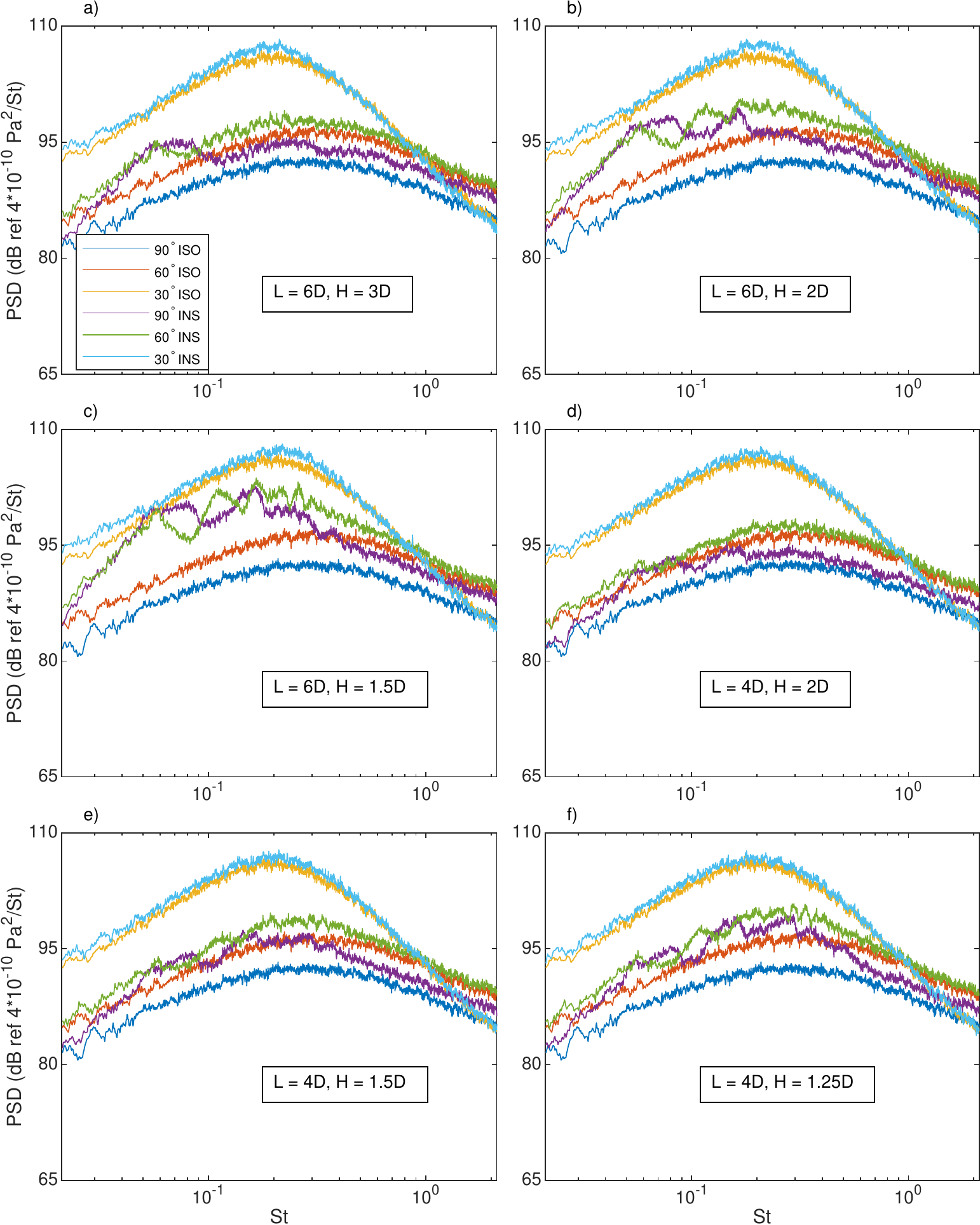}
    \caption{Isolated and installed noise spectra of a Mach number $0.7$ jet at
    $90^\circ$, $60^\circ$ and $30^\circ$ to the jet centreline on the
reflected side, for various plate positions shown inside each figure.}
    \label{fig:Reflected_RN01_ISOnINS_Mach07}
\end{figure*}

Experimentally measured installed jet noise spectra for $M_0 = 0.7$ are
presented in figures~\ref{fig:RN01_ISOnINS_Mach07} and
\ref{fig:Reflected_RN01_ISOnINS_Mach07}. Since jet noise power scales as the
eighth power of jet Mach number,\citep{Lighthill1952} one can see that
considerably larger noise spectra are observed compared to those shown in
figure~\ref{fig:RN01_ISOnINS_Mach05} and
\ref{fig:Reflected_RN01_ISOnINS_Mach05}. Nevertheless,
figure~\ref{fig:RN01_ISOnINS_Mach07} is qualitatively similar to
figure~\ref{fig:Reflected_RN01_ISOnINS_Mach07} in nearly all aspects. For
example, significant noise increase for installed jets occurs only at low
frequencies, placing the plate closer to the jet (decreasing the value of $H$
while $L$ is fixed) results in a stronger noise increase at low frequencies,
decreasing $L$ while $H$ is kept constant results in less significant noise
increase at low frequencies, etc. A more detailed discussion of these
characteristics is given at the end of the preceding section.


It is interesting to note the differences between the spectra at $M_0 = 0.5$
and $M_0 = 0.7$. One of the most striking differences is that the low-frequency
noise enhancement is less significant for the $M_0 = 0.7$ jets when the plate
is placed at the same positions. For instance, the maximum noise increase shown
in figure~\ref{fig:RN01_ISOnINS_Mach05}(c) is around $20$ dB while in
figure~\ref{fig:RN01_ISOnINS_Mach07}(c) around $15$ dB. The same conclusions
can be reached by comparing other pairs of spectra. This is believed to be due
to the dependence of the installed jet noise on jet Mach number. We have
mentioned in section~\ref{sec:introduction} that the intensity of installed jet
noise scales as the sixth power of the jet Mach number, while the isolated jet
noise intensity scales as the eighth power. Therefore, when the Mach number
increases, isolated jet noise power increases more quickly, hence narrowing the
relative difference between the installed and isolated noise spectra. This
trend is more evident by contrasting figures~\ref{fig:RN01_ISOnINS_Mach05}(f)
and \ref{fig:RN01_ISOnINS_Mach07}(f). 

The experimental results obtained in this section are consistent with earlier
findings. In addition, attention has also been paid to the effects of varying
$H$, $L$ and $M_0$ on the high-frequency installation behavior and the
non-symmetric dipolar directivity due to the refraction effects of the jet
plume. Full spectra at different observer angles for different $H$, $L$ and
$M_0$ combinations are presented in a detailed narrow-band manner. Such a
comprehensive database would be valuable in future studies.

\section{Comparison with the prediction model}
\label{sec:comparisonWithPredictionModel}
With a comprehensive set of experimental data on installed jet noise for
different plate positions available, we can compare them to the predictions of
the hybrid model briefly described in
section~\ref{sec:theHybridPredictionModel}. In particular, the innovative
instability-wave-scattering model can be properly validated. We chose to
compare the spectra at $M_0 = 0.5$. To do so, the PSD of the near-field
pressure fluctuations must be measured and used as an input to the model. For
convenience we also use the cylindrical coordinates centred around the jet exit
with $z$ in the streamwise, $r$ in the radial and $\varphi$ in the azimuthal
directions respectively.

\subsection{The near-field pressure}
\begin{figure*}[!hbtp]
    \centering
    \includegraphics[width=\widefigurewidth]{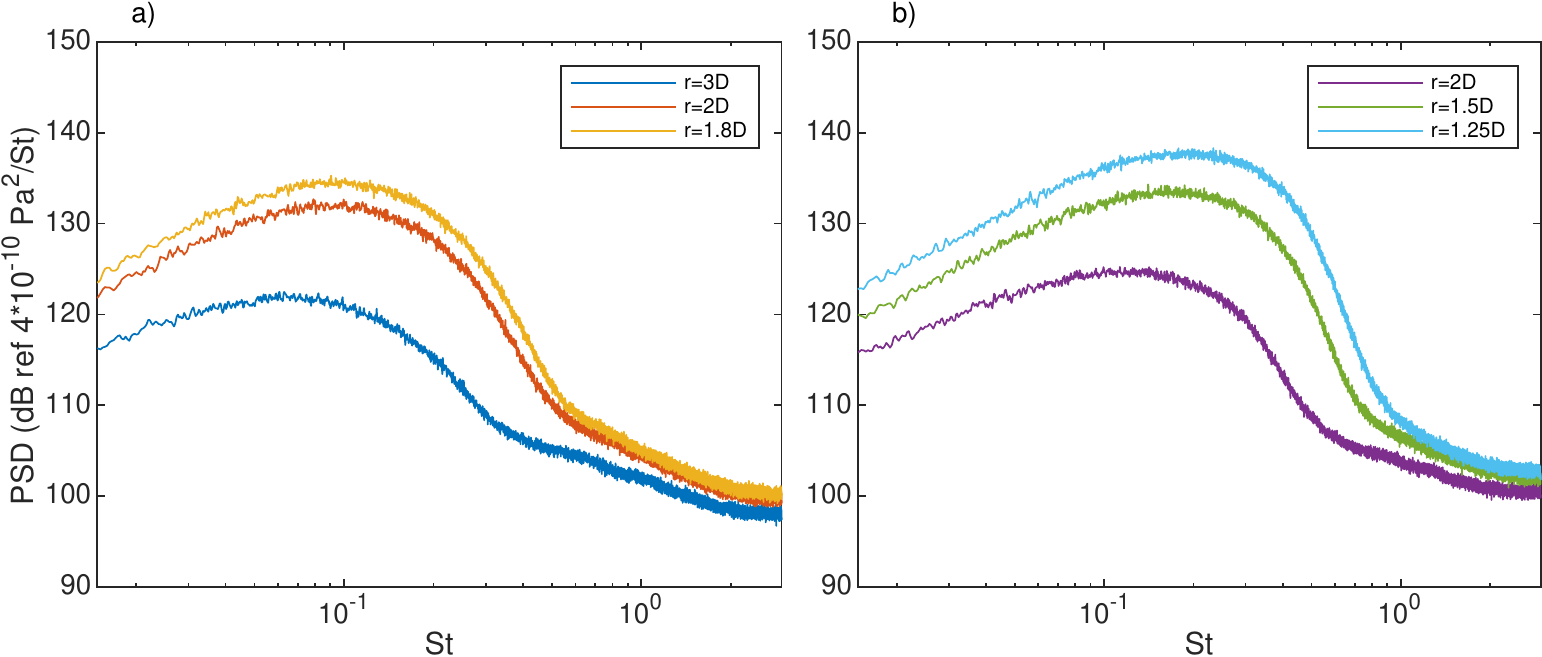}
    \caption{The power spectral densities of the near-field pressure at various
    radial positions but at two fixed streamwise locations: a) $z = 6D$; b) $z = 4D$. }
    \label{fig:nearFieldPSD}
\end{figure*}

Figure~\ref{fig:nearFieldPSD}(a) shows the power spectral densities of
the near-field pressure at $z = 6D$ but at various radial positions.
It should be noted that these spectra are obtained using one near-field
microphone (not to be confused with the far-field microphones) placed at a
specific azimuthal angle (not using a near-field azimuthal microphone arc).
Therefore, they are different from the modal pressure spectra mentioned in
section~\ref{sec:theHybridPredictionModel}. However, as noted in the earlier
paper,\citep{Lyu2016d} when the frequencies of interest are low, the sound due
to near-field scattering can be calculated by using the PSD of the near-field
pressure at a specific point. From figure~\ref{fig:nearFieldPSD}(a), it can be
seen that the near-field pressure fluctuation decays quickly as the radial
coordinate $r$ increases in the low and intermediate frequency regime. For
example, a decrease of up to $15$ dB is observed from $r= 2D$ to $3D$. However,
in the high-frequency regime, little change occurs. This is because the spectra
in the low-to-intermediate frequency range are dominated by the signature of
the instability waves, while acoustic perturbations dominate the near-field
pressure fluctuations at high frequencies, as demonstrated in the earlier
paper.~\citep{Lyu2016d} Also it can be observed that the spectral decay (due to
moving the microphone away from the jet) gradually increases as the frequency
increases, compatible with the behaviour of the modified Bessel functions of
the second kind (see \citet{Lyu2016d} for details). 

Figure~\ref{fig:nearFieldPSD}(b) shows the spectra at $z = 4D$. Comparing with
figure~\ref{fig:nearFieldPSD}(a), two distinctions can be readily observed.
Firstly, the spectra at $z = 4D$ but at the same radial position have smaller
amplitudes, especially at low frequencies. This is consistent with the
convecting growing behaviour of jet instability
waves.~\citep{Crow1971,Michalke1971} It is also consistent with the observation
that the installed jet noise is louder when the plate is placed further
downstream (but at the same radial positions, see the discussion on this in
section~\ref{sec:jetNoiseSpectra}). Secondly, spectra at $z = 4D$ have higher
peak frequencies. This is expected since it is known that instability waves at
high frequencies saturate earlier.~\citep{Crow1971,Michalke1984}

\begin{figure*}[!hbtp]
    \centering
    \includegraphics[width=\widefigurewidth]{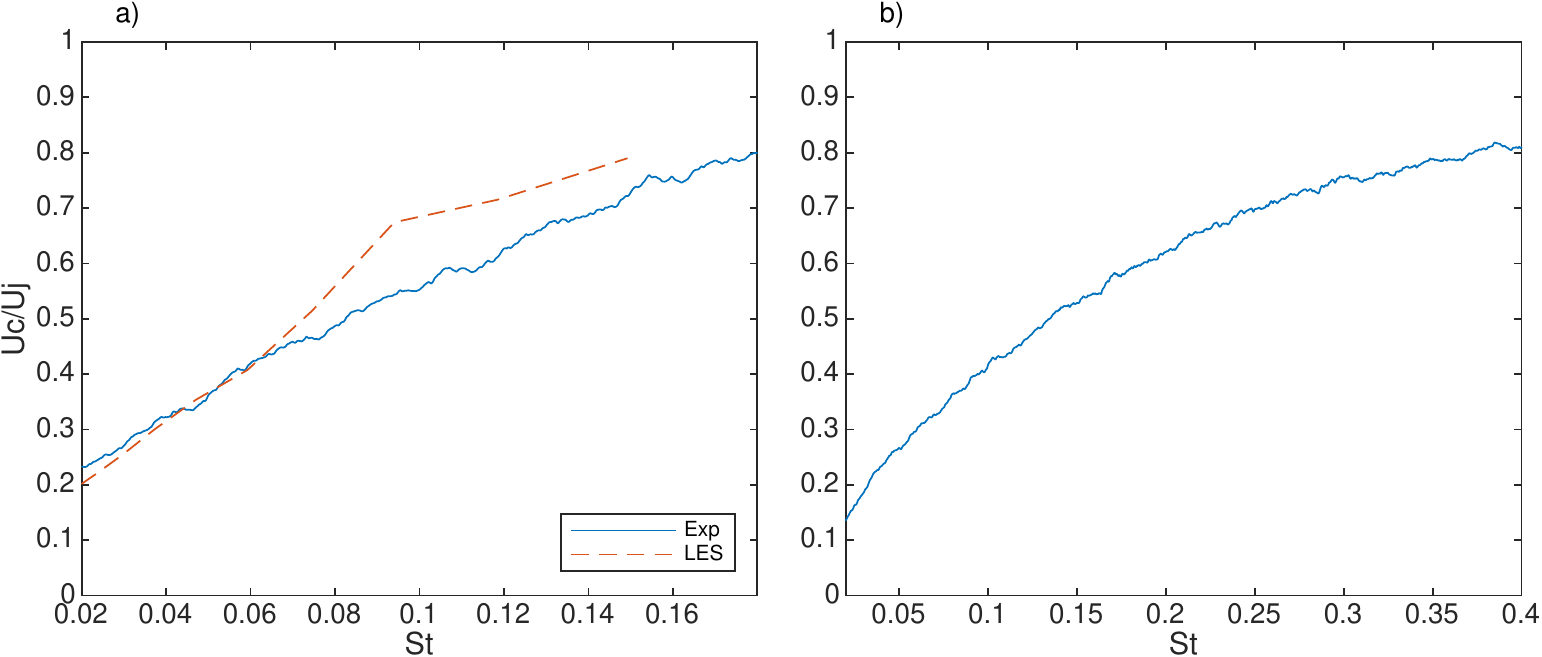}
    \caption{The convection velocities obtained from examining the radial
	decay rates of the near-field pressure PSDs at two different streamwise
    locations: a) $z = 6D$; b) $z = 4D$. }
    \label{fig:convectionVelocity}
\end{figure*}
The hybrid model described in section~\ref{sec:theHybridPredictionModel}
requires the spectral strengths and the frequency-dependent convection velocity
of the near-field instability waves as inputs. In the earlier
paper,\citep{Lyu2016d} this is calculated from analysing an LES database. A
similar analysis can be performed with the experimental data if we measure the
near-field pressure using a microphone arc, such as the one used by
\citetdot{Tinney2008b} But for simplicity, instead of using the modal spectra,
we use the overall spectra measured at one point. As discussed in the earlier
work,\citep{Lyu2016d} this is permissible, since the decay rates of the zero-
and first- order instability waves do not differ from each other significantly,
in particular at low frequencies. The convection velocity can also be estimated
by making use of this fact, i.e., we assume the point spectra decay at the same
rate as the zeroth-order instability waves. The convection velocities obtained
by examining the decay rates of the spectra shown in
figure~\ref{fig:nearFieldPSD} are plotted in
figure~\ref{fig:convectionVelocity}. Only low-frequency results are shown
because of the dominance of acoustic fluctuations at high frequencies. The
result for $z = 6D$ is shown and compared with that obtained from LES in
figure~\ref{fig:convectionVelocity}(a). Very good agreement is achieved. This
not only shows that the convection velocity is indeed frequency-dependent, but
also provides another piece of evidence for having an accurate LES database.
Figure~\ref{fig:convectionVelocity}(b) presents the convection velocity for the
instability waves at around $z = 4D$. Though the two curves in
figures~\ref{fig:convectionVelocity}(a) and \ref{fig:convectionVelocity}(b)
look qualitatively similar to each other, they are not strictly identical. This
is expected, since the mean flows at different streamwise locations are
different, so are the characteristics of jet instability waves.

\begin{figure*}[!hbtp]
    \centering
    \includegraphics[width=\widefigurewidth]{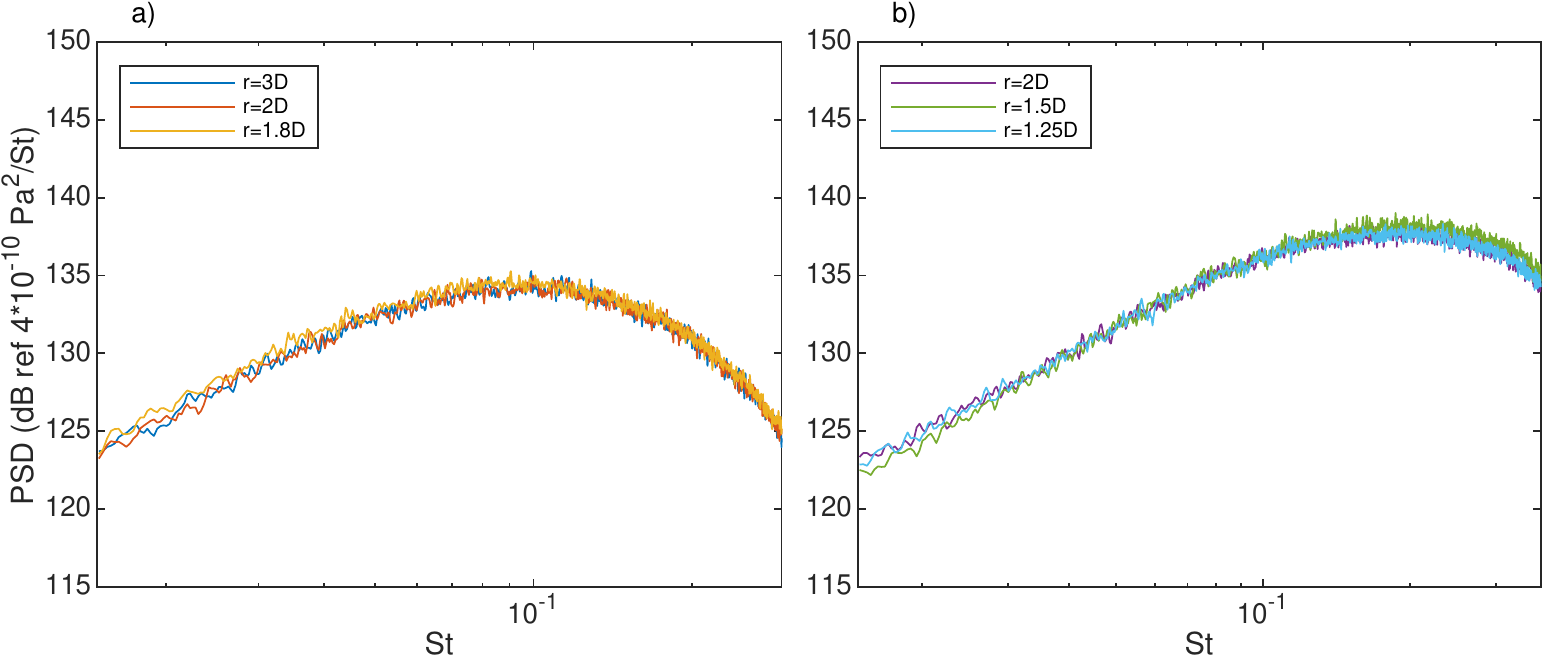}
    \caption{The collapsed power spectral densities of the near-field
    pressure using obtained convection velocities at various streamwise
locations: a) $z = 6D$; b) $z = 4D$.}
    \label{fig:collapsedNearFieldPSD}
\end{figure*}
Using the convection velocities obtained in
figure~\ref{fig:convectionVelocity}, the PSDs at various radial locations (but
the same streamwise location) can be successfully collapsed, as can be seen
from figure~\ref{fig:collapsedNearFieldPSD}. The plotted quantity in
figure~\ref{fig:collapsedNearFieldPSD} is the scaled PSD defined
by~\citep{Lyu2016d}
\begin{equation}
\frac{\Pi_0(\omega; r_0)}{K_0^2(\gamma_c r_0)},
    \label{equ:scaledNearFieldPSD}
\end{equation}
where $\Pi_0(\omega; r_0)$ denotes the overall PSD of the near field pressure
fluctuation measured at $r = r_0$. Note that the shown frequency range is
corresponding to the frequency range shown in
figure~\ref{fig:convectionVelocity}. Excellent agreement is achieved for
spectra at both $z = 6D$ and $z = 4D$, though there is slight disagreement near
the highest frequency for the spectra $z = 4D$. This is somewhat expected since
we assume different modal instability waves to decay at the same rate (at the
rate determined by the zeroth-order modified Bessel function of the second
kind). But high-order instability waves are more likely to exist at upstream
locations.\citep{Tinney2008a}

\subsection{Validation of the hybrid prediction model}
Equipped with the flow data from similar RANS simulations to that described in
the earlier paper,\citep{Lyu2016d} the frequency-dependent convection velocity
and the near-field pressure PSDs from experiment, the hybrid model can be
evaluated readily. The near-field pressure PSD measured at the smallest value
of $r_0$ is used as the input for each value of $L$. Using those measured at
other values leads to negligible change to the far-field predictions. This is
because the instability waves decay with radial distance is accurately
predicted with the use of the frequency-dependent convection velocity. The
predicted sound spectra are compared with experimentally measured PSDs at
various observer angles and for various plate positions. In the rest of this
section, comparisons are shown for six observer angles for each plate position,
i.e., $90^\circ$, $60^\circ$ and $30^\circ$ to the jet axis on both the
shielded and reflected sides of the flat plate. 

\subsubsection{For plate position at \texorpdfstring{$L = 6D$}{L=6D}}
\begin{figure*}[!htbp]
  \centering 
  \includegraphics[width=\widefigurewidth]{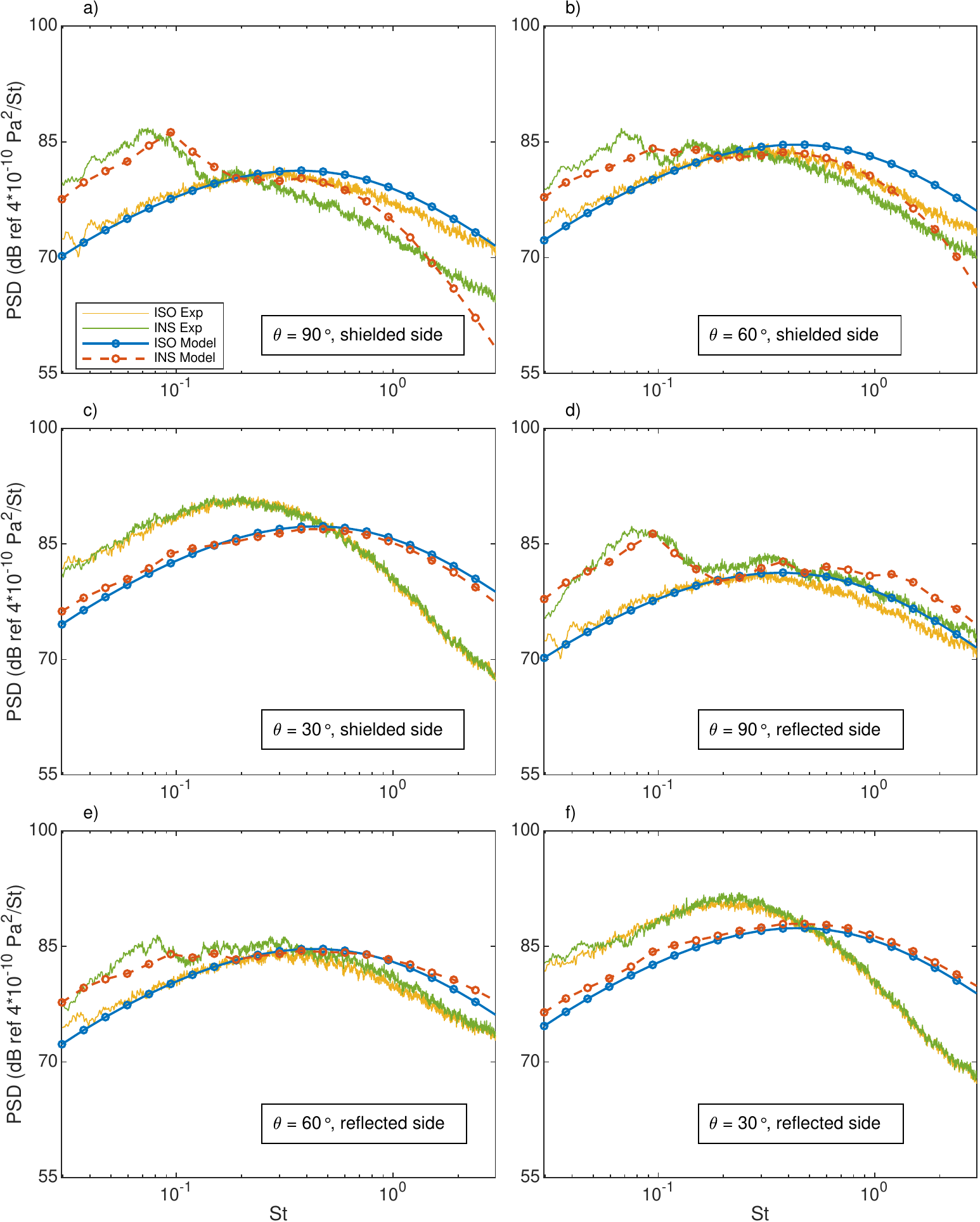}
  \caption{Comparison of the predicted isolated and installed noise spectra
  with experimental results at various observer locations for $L = 6D$ and $H =
  3D$.}
    \label{fig:valModel_RN01_X6R3_Mach05}
\end{figure*}
Figure~\ref{fig:valModel_RN01_X6R3_Mach05} shows the comparison of the isolated
and installed jet noise spectra when the plate's trailing edge is at $L = 6D$
and $H = 3D$. We should emphasise here that the isolated jet noise spectra are
predicted using the Lighthill's acoustic analogy described in
section~\ref{sec:theHybridPredictionModel} with a free-field Green's function,
and so neglect the refraction effects by the mean flow. We discuss the results
on the shielded side first (figure~\ref{fig:valModel_RN01_X6R3_Mach05}(a-c)).
The agreement between the isolated spectra at $\theta = 90^\circ$
(figure~\ref{fig:valModel_RN01_X6R3_Mach05}(a)) is very good, apart from a
slight over-prediction at high frequencies. The installed spectra are much more
interesting. At such a distance to the jet, the maximum noise enhancement at
low frequencies caused by the scattering of instability waves is around $10$
dB. This is well captured by the model. At high frequencies, noise is
effectively shielded by the plate. This is qualitatively captured by the
prediction model (though the predicted amplitude is not very different from the
experimental results as well). This discrepancy may be explained by the fact
that jet refraction effects are not included in the Lighthill acoustic analogy
part of the hybrid model.\citep{Lyu2016d}
Figure~\ref{fig:valModel_RN01_X6R3_Mach05}(b) shows the comparison for $\theta
= 60^\circ$. The prediction for the isolated spectrum starts to deviate from
the experimental result at high frequencies due to the neglect of jet
refraction effects, though at low frequencies the agreement is still
acceptable. For the installed noise spectra, as discussed in the preceding
section, weaker noise enhancement and shielding effects are observed
experimentally. The model can successfully capture this and agrees with the
experimental results at low frequencies. The trend of weaker shielding effects
is also exhibited by the model. The spectra comparison at $\theta = 30^\circ$
is more interesting. One can see that at such a low observer angle, jet
refraction is significant, even for low frequencies. Therefore, the deviation
between the experiment and the prediction for isolated noise spectra can be as
large as $5$ dB. However, the fact that there is little noise enhancement at
low frequencies and little noise suppression at high frequencies is captured
remarkably well. In other words, the model successfully captures the physics of
jet installation, although the absolute values of the predicted spectra are
affected by the refraction effects. This indicates that apart from refraction
effects, the agreement for installed jet noise spectra is very good.

Noise comparison on the reflected side is shown in
figure~\ref{fig:valModel_RN01_X6R3_Mach05}(d-f). Since the isolated jet noise
spectra are identical to those on the other side of the plate, we only need to
focus on the installed spectra. Figure~\ref{fig:valModel_RN01_X6R3_Mach05}(d)
shows the results at $90^\circ$ to the jet axis. Excellent agreement at this
observer angle is achieved since there are no (little) refraction effects. At
low frequencies, the prediction is identical (symmetric) to that on the other
side of the plate and the agreement continues to be good. At high frequencies,
the noise increase of around $3$ dB is successfully predicted by the model. The
slight over-prediction for the high-frequency installed spectrum is apparently
caused by the slight over-prediction of the isolated spectrum. Comparison of
noise spectra at $60^\circ$ to jet axis is shown in
figure~\ref{fig:valModel_RN01_X6R3_Mach05}(e). As discussed above, the same
level of agreement as on the other side of plate is achieved at low
frequencies. However, we see that the predicted spectrum gradually deviates
from the experimental spectrum at high frequencies due to the refraction
effects. The tendency of weaker noise increase due to reflection at lower
observer angles is captured. Figure~\ref{fig:valModel_RN01_X6R3_Mach05}(f)
shows the comparison at $30^\circ$ to the jet axis. This is very similar to the
results on the other side of the plate, for the same reasons.

\begin{figure*}[!htbp]
  \centering
    \includegraphics[width=\widefigurewidth]{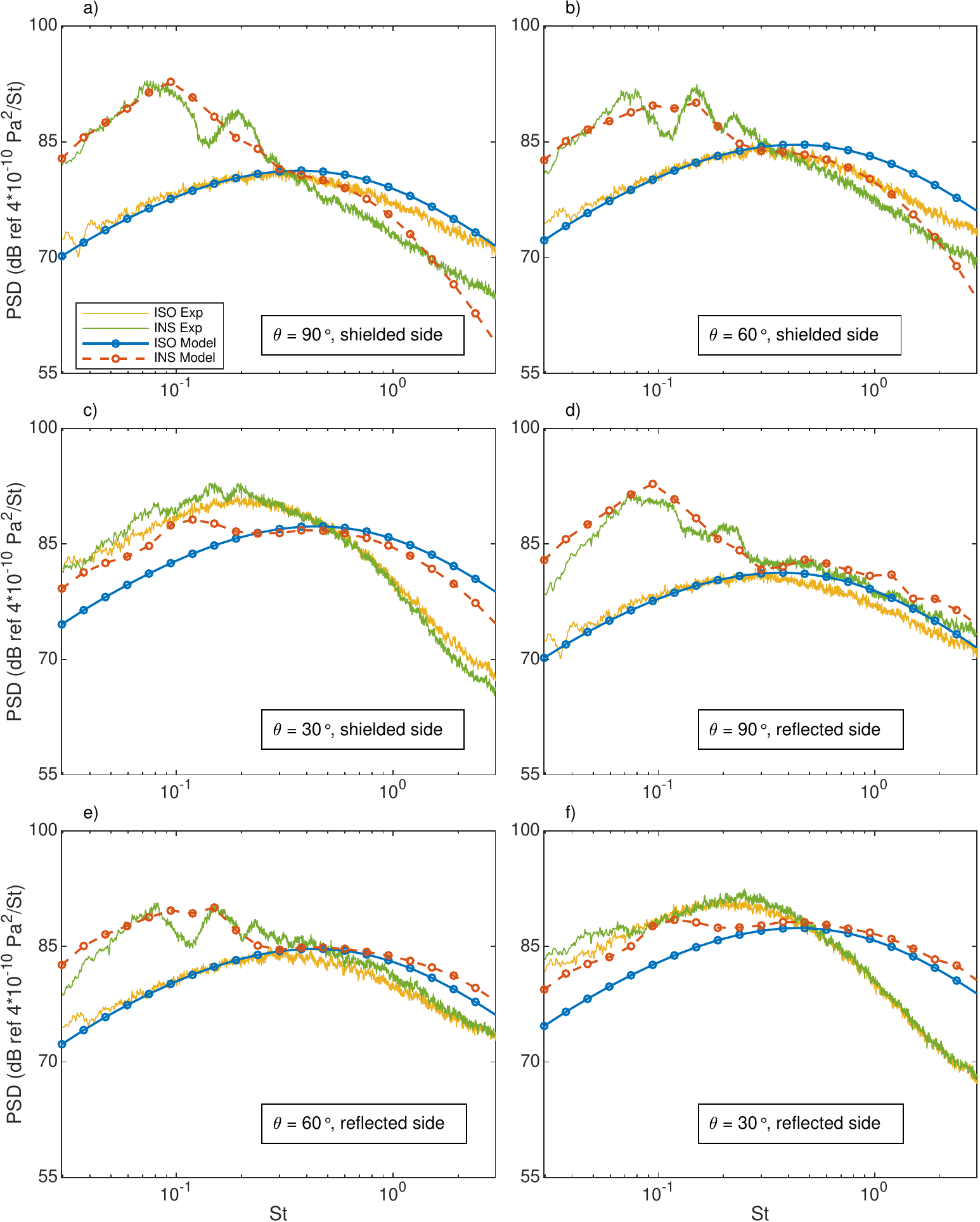}
    \caption{Comparison of the predicted isolated and installed noise spectra
    with experimental results at various observer locations for $L = 6D$ and $H
= 2D$.}
    \label{fig:valModel_RN01_X6R2_Mach05}
\end{figure*}
\begin{figure*}[!htbp]
  \centering
    \includegraphics[width=\widefigurewidth]{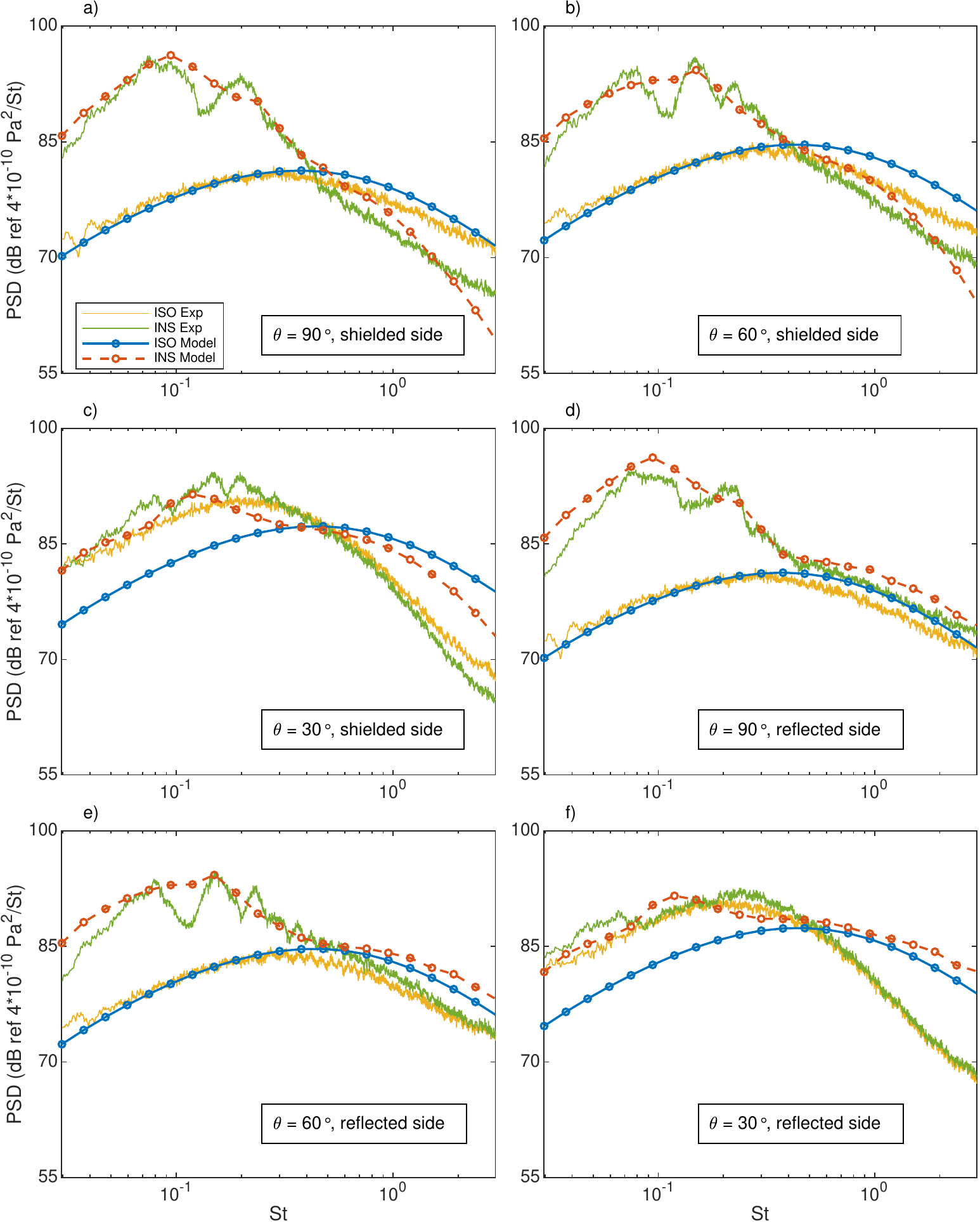}
    \caption{Comparison of the predicted isolated and installed noise spectra
    with experimental results at various observer locations for $L = 6D$ and $H
= 1.5D$.}
    \label{fig:valModel_RN01_X6R1p5_Mach05}
\end{figure*}
Comparison of the spectra when the plate's trailing edge is at $L = 6D$ and $H
= 2D$ is shown in figure~\ref{fig:valModel_RN01_X6R2_Mach05}. Again, identical
isolated jet noise spectra suggest that we only need to discuss installed
spectra. Figure~\ref{fig:valModel_RN01_X6R2_Mach05}(a) shows excellent
agreement between the experimental and predicted spectra at $\theta =
90^\circ$. The low-frequency noise intensification is much more pronounced at
such a plate position and is predicted remarkably well. Even the high-frequency
shielding effects are predicted reasonably well. Note that there are small
oscillations in the experimental spectrum, which are not captured very well by
the model. One can show that this oscillation is due to the finite chord length
of the flat plate (see more details in Appendix A). The model makes use many
approximations in order to reach a simplified formula, which might be the
reason for not capturing these low-frequency oscillations. However, the
predicted spectrum does follow the mean experimental counterpart very closely.
Figure~\ref{fig:valModel_RN01_X6R2_Mach05}(b) shows the results for an observer
at $\theta = 60^\circ$. Excellent agreement continues to be achieved at low
frequencies, while the high-frequency prediction sees discrepancies due to
refraction effects. Figure~\ref{fig:valModel_RN01_X6R2_Mach05}(c) deserves a
detailed explanation. At first glance it appears that the prediction yields a
much higher noise enhancement at low frequencies, which does not agree with
experimental observations. However, one should bear in mind that the total
power spectral density $\Phi$ is the sum of $\Phi_Q$ and $\Phi_N$. Therefore, a
lower value of $\Phi_Q$ (which is exhibited by the much lower value of the
isolated spectrum in the figure and the fact that the isolated spectrum is
nearly identical to the installed spectrum at low frequencies) would contribute
little to $\Phi$ and the total spectrum $\Phi$ is nearly solely determined by
the large value of $\Phi_N$. Hence $\Phi$ would be much larger than $\Phi_Q$
(hence the isolated spectrum, which explains the significant noise augmentation
predicted in figure~\ref{fig:valModel_RN01_X6R2_Mach05}(c)). However, if
$\Phi_Q$ had been correctly predicted to be of larger values, the contribution
from $\Phi_N$ would have been much less pronounced, and one would have expected
only a slight noise increase. Therefore, the seemingly discrepancy at low
frequencies is due to the inaccurate prediction of the isolated jet noise
spectrum and the near-field scattering model works remarkably well (we can see
evidence of this if we add the $\Phi_N$ to the isolated spectra measured in the
experiment).

The comparison on the reflected side is shown in
figure~\ref{fig:valModel_RN01_X6R2_Mach05}(d-f). Excellent agreement is
observed at $90^\circ$ to the jet axis (see
figure~\ref{fig:valModel_RN01_X6R2_Mach05}(d)). The less good agreement at the
lowest frequencies is caused, as mentioned in the preceding section, by the
refraction of the reflected sound by the jet, which is not accounted for in
this model. High-frequency agreement is nearly identical to that shown in
figure~\ref{fig:valModel_RN01_X6R3_Mach05}(d).
Figure~\ref{fig:valModel_RN01_X6R2_Mach05}(e) shows the results when the observer is
at $60^\circ$ to the jet axis. Apart from the larger noise intensification,
which is correctly captured by the model, it is again similar to the comparison
shown in figure~\ref{fig:valModel_RN01_X6R3_Mach05}(e). So are the results
shown in figure~\ref{fig:valModel_RN01_X6R2_Mach05}(f).

When the plate is moved closer to the jet at $H =1.5D$, a significant noise
increase of up to $20$ dB is achieved. The comparison of the model prediction
with experimental results at such a close distance is shown in
figure~\ref{fig:valModel_RN01_X6R1p5_Mach05}. The hybrid model, especially the
instability-wave-scattering model predicts the noise enhancements at all angles
remarkably well. Note again that the seemingly over-prediction of the
low-frequency enhancement in figures~\ref{fig:valModel_RN01_X6R1p5_Mach05}(c)
and \ref{fig:valModel_RN01_X6R1p5_Mach05}(f) is caused by the inaccurate
prediction of the isolated spectra. 

\subsubsection{For plate position at \texorpdfstring{$L = 4D$}{L=4D}}
\begin{figure*}[!htbp]
  \centering
    \includegraphics[width=\widefigurewidth]{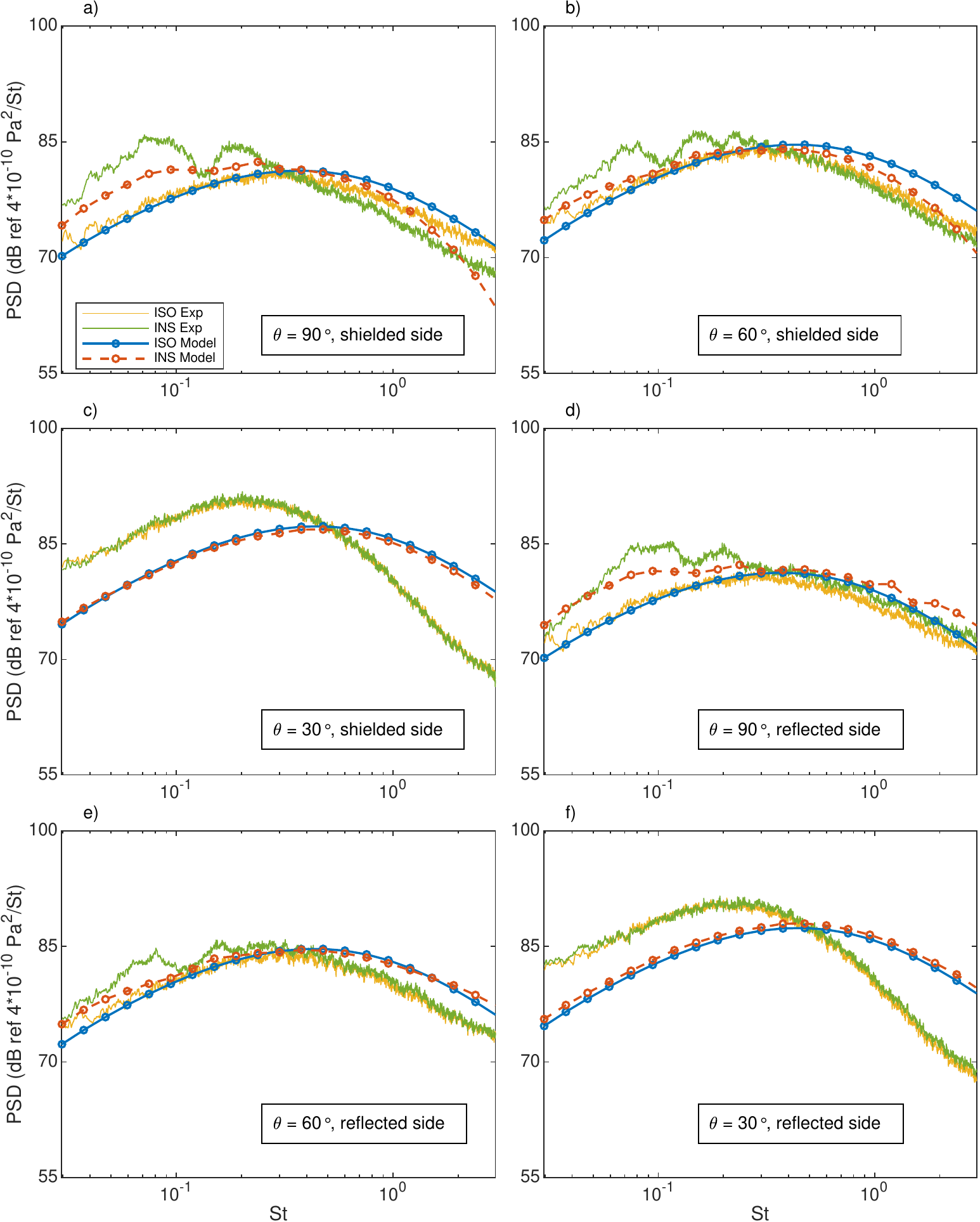}
    \caption{Comparison of the predicted isolated and installed noise spectra
    with experimental results at various observer locations for $L = 4D$ and $H = 2D$.}
    \label{fig:valModel_RN01_X4R2_Mach05}
\end{figure*}

\begin{figure*}[!htbp]
  \centering
    \includegraphics[width=\widefigurewidth]{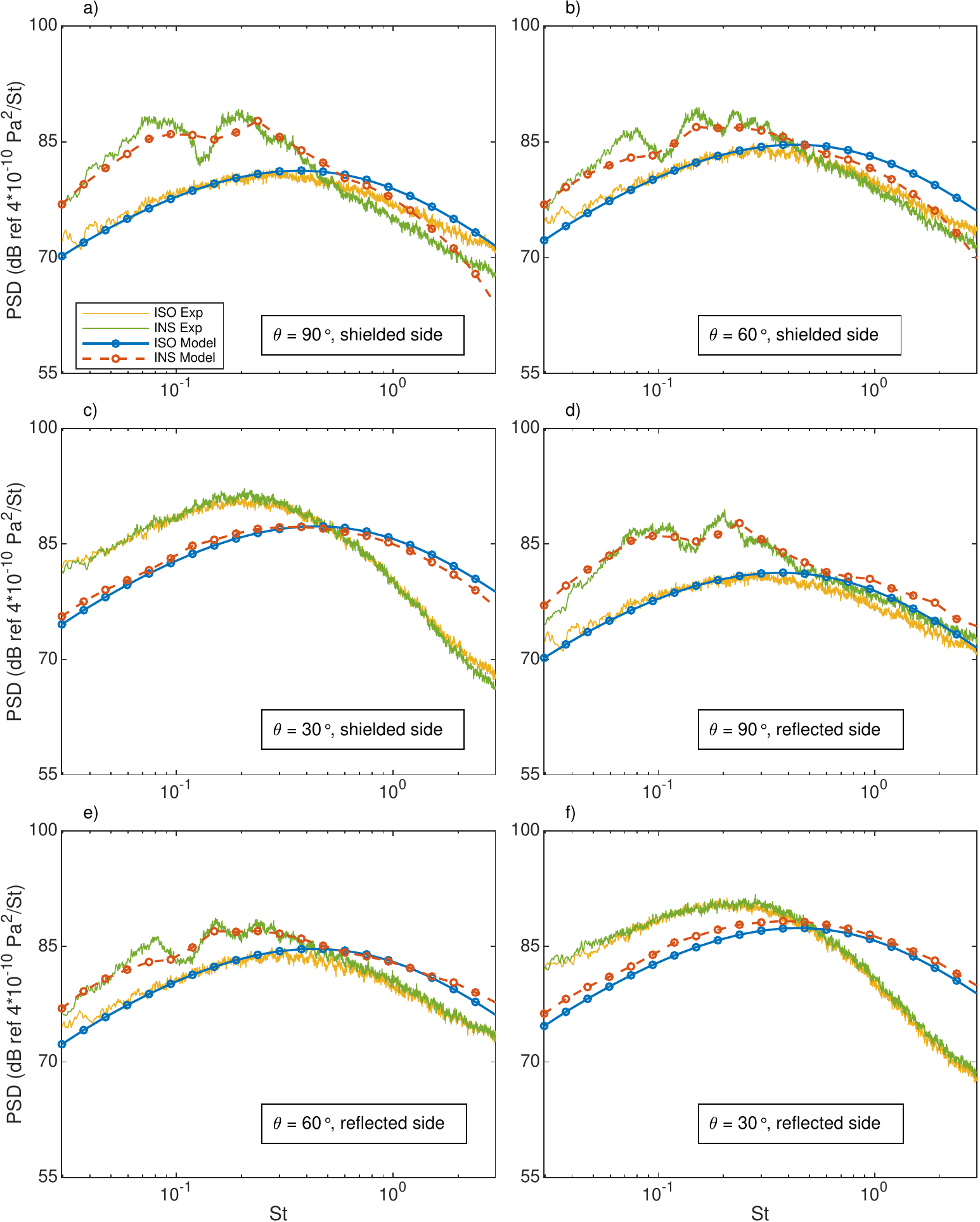}
    \caption{Comparison of the predicted isolated and installed noise spectra
    with experimental results at various observer locations for $L = 4D$ and $H
= 1.5D$.}
    \label{fig:valModel_RN01_X4R1p5_Mach05}
\end{figure*}
Comparison of the isolated and installed jet noise spectra when $L = 4D$ and $H
= 2D$ is shown in figure~\ref{fig:valModel_RN01_X4R2_Mach05}. As already noted
in the section discussing the experimental results, moving the plate towards
jet nozzle causes the noise increase at low frequencies to be less significant.
Results are again shown for both the shielded
(figure~\ref{fig:valModel_RN01_X4R2_Mach05}(a-c)) and reflected sides
(figure~\ref{fig:valModel_RN01_X4R2_Mach05}(d-f)). At $\theta = 90^\circ$ on
the shielded side, a noise increase of up to $8$ dB is found. The prediction
yields a noise increase of around $5$ dB, slightly below the experimental
results. It is suggested that this could be either due to the spectral
oscillations or because the approximation that the instability waves of
different azimuthal modes decay at roughly the same rate is less good at $z =
4D$ (because the instability waves of higher azimuthal modes do not vanish as
quickly as those at $z = 6D$, see \citet{Tinney2008a} and \citet{Tinney2008b}
for instance). Comparison at $\theta = 60^\circ$ shows similar level of
agreement as that at $\theta = 90^\circ$, and the spectra plotted in
figure~\ref{fig:valModel_RN01_X4R2_Mach05}(c) resemble those shown in
figure~\ref{fig:valModel_RN01_X6R3_Mach05}(c). Results on the other side of the
plate are shown in figure~\ref{fig:valModel_RN01_X4R2_Mach05}(d-f). Due to the
similarity to those discussed above, a detailed discussion seems superfluous.

It is, however, worth noting that the agreement between the model predictions
and experimental observations is much better when the plate is moved slightly
closer to $H = 1.5D$, as shown in figure~\ref{fig:valModel_RN01_X4R1p5_Mach05}.
The maximum noise enhancement observed at $90^\circ$ on the shielded side in
the experiment matches closely to the model's prediction.
\begin{figure*}[!htbp]
  \centering
    \includegraphics[width=\widefigurewidth]{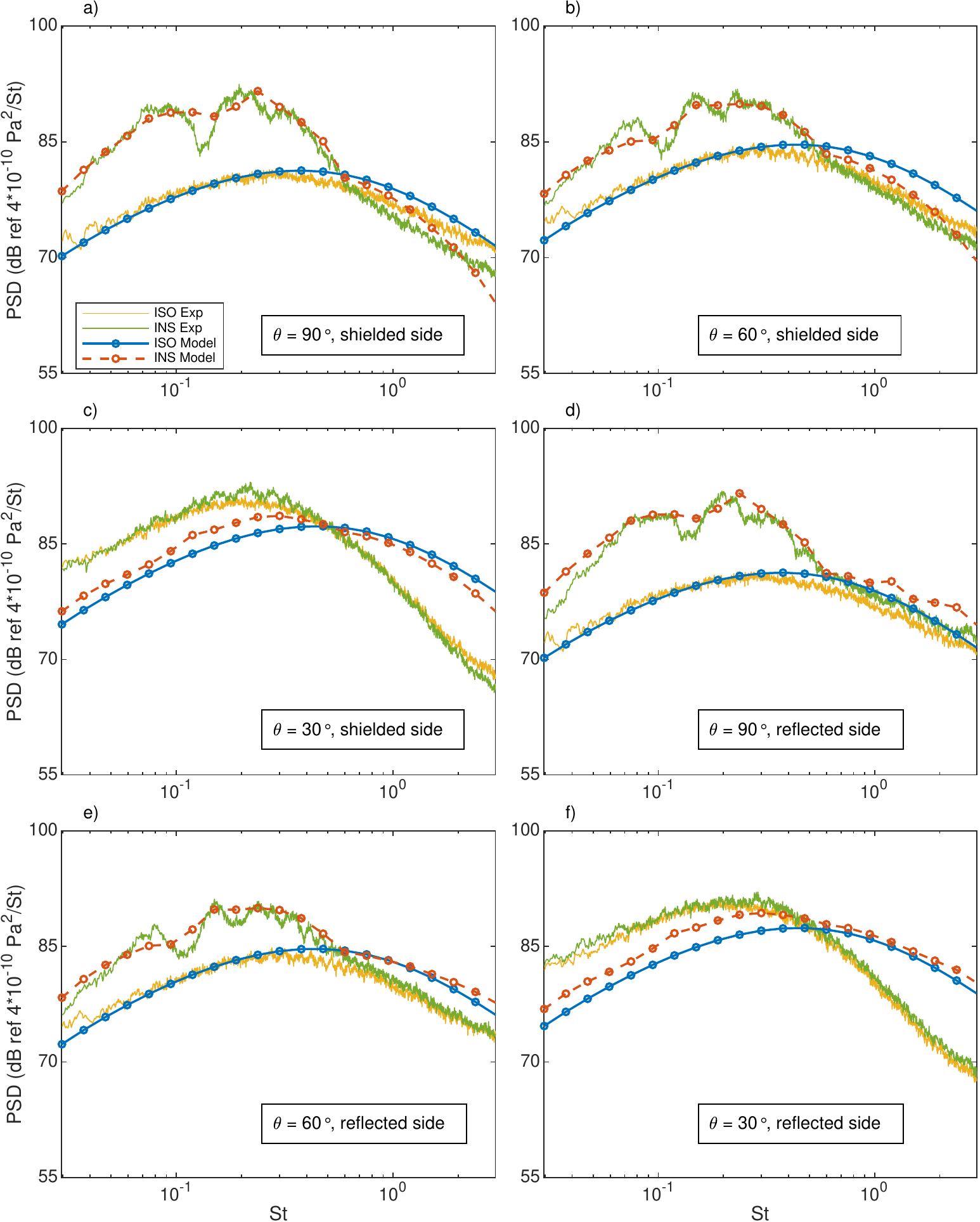}
    \caption{Comparison of the predicted isolated and installed noise spectra
    with experimental results at various observer locations for $L = 4D$ and
$H =1.25D$.}
    \label{fig:valModel_RN01_X4R1p25_Mach05}
\end{figure*}
Even the spectral oscillations appear to be partially predicted. Noise
shielding effects for this configuration are also predicted reasonably well. The
agreement at both $60^\circ$ and $30^\circ$ to the jet axis is similar to that
described above. And it is worth mentioning again the excellent noise
prediction at $90^\circ$ on the reflected side.

One can further compare the results when the plate's trailing-edge is placed at 
$L = 4D$ and $H = 1.25D$. The fact that the spectral oscillations are partially
captured is more marked. The general trend, however, largely resembles that
shown in figure~\ref{fig:valModel_RN01_X4R1p5_Mach05}. We therefore omit a
repetitive description of them.

\section{Conclusion}
A series of experimental tests are carried out in this paper to investigate jet
installation effects, together with the effects of varying $H$, $L$ and the jet
Mach number on installed jet noise. It is found that the plate causes jet noise
to be enhanced significantly at low frequencies, and jet noise is either
suppressed or increased by around $3$ dB at high frequencies on the shielded
and reflected sides, respectively. It is demonstrated that increasing $H$
(while $L$ is fixed) causes the low-frequency amplification to decrease
exponentially but results in little change for both the shielding and
reflection effects at high frequencies. Increasing $L$ (while $H$ is fixed), on
the other hand, produces stronger noise intensification at low frequencies and
slightly more effective shielding or reflection effects at high frequencies.
The installation effects are found to be less pronounced as the jet Mach
number increases.

The results are then compared with the predictions using the hybrid model
developed in the earlier work of the authors.\citep{Lyu2016d} Excellent
agreement is achieved for the low-frequency noise enhancement caused by
instability wave scattering when the plate is placed at different positions.
This remarkable agreement shows that the near-field scattering model captures
the correct noise mechanism for the low-frequency noise enhancement and
provides a robust and accurate prediction tool for the installation effects for
a given isolated jet. In addition, the quadrupole-scattering model can also
correctly predict the noise spectra for an observer angle of $90^\circ$ on the
reflected side of the flat plate. At lower observer angles, deviations occur
due to the jet refraction effects. However, it can qualitatively predict both
the shielding and reflection effects at high frequencies. An improved model
incorporating these refraction effects will be studied in our future work. In
addition, the frequency dependence of the instability waves' convection
velocity in the low-frequency regime appears to be very interesting. This will
be studied from a stability analysis perspective in future work.

%
%

%

\begin{acknowledgments}
The first author (B. Lyu) wishes to gratefully acknowledge the financial 
support provided by the Cambridge Commonweath European and Internatinoal Trust 
and the China Scholarship Council.
\end{acknowledgments}

\section*{Appendix}
\subsection{On the spectral oscillations} 
To show that the oscillations observed in the installed jet noise spectra are
due to the finite chord length of the flat plate, first note that the
frequencies where the oscillatory peaks and troughs appear remain unchanged
when either $L$ or $H$ changes. This can be seen, for example, by comparing
figures 15a, 16a, 18a, and 19a. This shows that the cause of such an
oscillation is most likely to be related with the size of the plate, which is
kept constant.

The existing literature on the turbulent boundary layer trailing-edge noise (TE
noise) can be used to support such an argument quantitatively (see, for
example, \citetcomma{Amiet1976b} \citetcomma{Roger2005} and \citet{Lyu2016}).
TE noise is a common issue in applications such as wind turbines and cooling
fans. It refers to the noise generated when the turbulence in the boundary
layer convects past and gets scattered by the sharp trailing edge of an
aerofoil. Amiet's approach~\citep{Amiet1976b} is widely used in the
aeroacoustic community to model the TE noise. Amiet's model was developed in
two steps. Firstly, the aerofoil was simplified as a flat plate, and the
scattered pressure on the upper and lower surfaces was calculated using the
Schwarzschild method. In doing so, the leading edge of the aerofoil was often
assumed to be far away from the trailing edge and hence cause minimal back
scattering. The work of \citet{Roger2005} extended Amiet's model by taking into
account the leading-edge back scattering. It found that when $kc > 1$ the
leading-edge back scattering is negligible. Secondly, the far-field sound was
obtained by integrating the surface pressure based on the theory of Kirchhoff
and Curle.~\citep{Curle1955} Hence, we can see that the finite size of the
plate was partially taken into account (no leading-edge back scattering, but
the surface pressure integral domain is finite). Because the chord length is
finite, one expects some oscillatory pattern in the spectrum because the
surface pressure has an oscillatory pattern along the
chord.\citep{Amiet1975,Amiet1976b} Figure 7a (black dashed line) in the work of
\citet{Lyu2016} represents a typical TE noise spectrum. The oscillatory nature
is clear.
 
The instability-wave scattering model shown in
Section~\ref{sec:theHybridPredictionModel} uses Amiet's approach to obtain the
far-field sound. The oscillatory nature would also appear in the installed jet
noise context. We can make a quantitative comparison between locations where
the first trough and peak appear between the aforementioned figure and those
observed in the experiment. It has been shown that the frequency where the peak
or trough appears has a strong dependence on $k$ and a weak dependence on $k_1$
(see \citetcomma{Roger2005} and \citet{Lyu2016}). Therefore, though the values
of $k_1$ are different between the aforementioned Figure~7a and our experiment,
we expect the peak or tough frequency to more or less match each other. The TE
noise figure shows that the first dip frequency at around $kc=5.2$ and the
following peak frequency at around $kc =8$. In the experiment $c = 12D$ and
these correspond to the frequencies at $931\textrm{ Hz}$ ($St \approx 0.14$)
and $1433\textrm{ Hz}$ ($St \approx 0.21$), respectively. This matches the
experimental results very well. We therefore believe that this oscillatory
pattern is due to the finite chord length of the flat plate. 

\bibliography{cleanRef}
\end{document}